\documentclass[12pt]{article}
\usepackage[T1]{fontenc} 
\usepackage{breakcites} 
\usepackage{url} 

\newcommand{\itemb}{\begin{itemize}}
\newcommand{\iteme}{\end{itemize}}

\newcommand{\enumb}{\begin{enumerate}}
\newcommand{\enume}{\end{enumerate}}

\newcommand{\refsection}[2]{\section{#1}\label{sec:#2}}

\newcommand{\smallest}[1]{\vspace{6pt}\noindent {\bf  {#1}.}}

\newcommand{\com}[1]{}
\usepackage{xcolor}

\newcommand{\gx}{glucocorticoid}

\newcommand{\testos}{testosterone}

\newcommand{\sz}{schizophrenia}
\newcommand{\Sz}{Schizophrenia}

\newcommand{\epil}{epilepsy}

\newcommand{\pufa}{polyunsaturated fatty acid}

\newcommand{\PUFA}{Polyunsaturated Fatty Acid}

\newcommand{\pla}{phospholipase A}

\newcommand{\indep}{independent}

\newcommand{\spont}{spontaneous}

\newcommand{\adol}{adolescence}

\newcommand{\ecb}{endocannabinoid}
\newcommand{\Ecb}{Endocannabinoid}
\newcommand{\antip}{antipsychotic}
\newcommand{\Antip}{Antipsychotic}

\newcommand{\er}{endoplasmic reticulum}

\newcommand{\antag}{antagonist}

\newcommand{\xtblty}{excitability}

\newcommand{\epig}{epigenetic}

\newcommand{\sensit}{sensitization}

\newcommand{\desensit}{desensitization}

\newcommand{\metab}{metabolism}

\newcommand{\amg}{amygdala}

\newcommand{\hippo}{hippocampus}

\newcommand{\oxis}{{{oxidative stress}}}

\newcommand{\icell}{{{intracellular}}}

\newcommand{\ecell}{{{extracellular}}}

\newcommand{\transm}{{{transmission}}}

\newcommand{\homeos}{{{homeostasis}}}

\newcommand{\degen}{{{degeneration}}}

\newcommand{\infl}{{{inflammation}}}

\newcommand{\astc}{astrocyte}

\newcommand{\plast}{{{plasticity}}}
\newcommand{\Plast}{{{Plasticity}}}
\newcommand{\mitoch}{{{mitochondria}}}

\newcommand{\catwo}{{{Ca}\textsuperscript{{\raisebox{-1pt}{$\scriptstyle{2+}$}}}}}

\newcommand{\kplus}{{{K}\textsuperscript{+}}}

\usepackage[superscript]{cite} 
\usepackage{changepage} 
\usepackage{times}
\usepackage{cite} 
\usepackage{hyperref}
\usepackage[margin=0.5cm]{caption} 

\newcommand{\thn}{P*SZ} 

\usepackage{lineno} 
\usepackage{fancyhdr}
\begin{document}
%
\title{A \PUFA\ (PUFA)\\ Theory of \Sz}
%
\author{
{Ari Rappoport}\\ The Hebrew University of Jerusalem\\{ari.rappoport@mail.huji.ac.il}\\
}
\date{Summer 2023}
\maketitle
\begin{adjustwidth}{30pt}{30pt}
{\bf Abstract.} 
I present a theory of \sz\ (SZ) that mechanistically explains its etiology, symptoms, pathophysiology, and treatment. SZ involves the chronic release of membrane \pufa s (PUFAs) 
and their utilization for the synthesis of stress-induced \plast\ agents such as \ecb s (ECBs). 
The causal event in SZ is prolonged stress during a sensitive period, which can induce prolonged and heritable changes. 
The physiological effect of the released PUFAs and their products is to disconnect neurons from their inputs and promote intrinsic \xtblty. 
I show that these effects
can explain the positive, negative, cognitive, and mood symptoms of SZ, as well as the mechanisms of many known triggers of psychosis. 
The theory is supported by overwhelming evidence addressing lipids, immunity, ECBs, neuromodulators, hormones, neurotransmitters, and cortical parameters in SZ. 
It explains why \antip\ drugs are effective against positive symptoms, and why they do not affect the other symptoms.  
Finally, I present promising treatment directions implied by the theory, including some that are immediately available. 
 
\end{adjustwidth}

\refsection{Introduction}{intro}
\Sz\ (SZ) is one of the truly major psychiatric disorders, involving hallucinations, delusions (positive symptoms), reduced motivation, activity, speech, affect, sociality, and pleasure (negative symptoms), impaired cognition, and mood symptoms \cite{tandon2008sch}. 
It has a strong heritable component. 
SZ is typically diagnosed at late adolescence, with prodromal non-positive symptoms. It affects dozens of millions of people, and has a strong devastating impact throughout the person's life. Treatment is generally effective only against positive symptoms. Suicide and relatively early death are common. 


At present, there is no theory that explains SZ. Several hypotheses have been formulated \cite{fivsar2022bio}, 
but each of them addresses only a partial subset of the data, and none of them provides a mechanistic account of the disorder. The standard hypothesis, motivated by the effects of dopamine (DA) agonists and \antag s on psychosis, is that SZ is caused by DA imbalance. However, this hypothesis is neither fully explanatory nor supported by strong evidence \cite{owen2016sz}.
\Ecb s (ECBs) \cite{leweke2017put} and \pufa s (PUFAs) \cite{hoen2013red} are thought to be involved in SZ, but there is no specific proposal for exactly how. 

This paper presents the {\bf PUFA Theory of \Sz\ (\thn)}, a theory of SZ that addresses all of the salient data known about the disorder. Briefly, \thn\ posits that {\em \sz\ is caused by chronic synthesis and signaling of PUFAs and PUFA products associated with stress-induced \plast, including ECBs, due to stress experienced at sensitive periods that yields persistent and heritable changes. The chronic presence of these processes can mechanistically explain the full range of SZ symptoms and data}. 

\thn\ is conceptually simple, 
differing from previous SZ hypotheses in two major ways. First, it is a complete theory, not a hypothesis, providing a detailed mechanistic account of the disorder, including its etiology, symptoms, pathophysiology, and treatment. Second, it unifies virtually all of the existing SZ evidence into a single coherent story. This includes the two largest bodies of evidence in SZ, lipids and ECBs, as well as evidence related to immunity, pain, neuromodulators, hormones, neurotransmitters, symptoms, and cortical parameters. 

Below, I present the theory, explain the mechanisms and roles of PUFAs and ECBs in health and stress,  
and detail extensive evidence supporting the theory. I then discuss possible treatments implied by the theory, some of which not requiring new drug development.

\refsection{PUFA Theory of \Sz\ (\thn )}{theory}
\smallest{Motivation}
Examination of strong risk factors and their inter-relationships is a good starting point when trying to understand a disorder. 
The following four pieces of data are well-known and supported by overwhelming evidence (further detailed below).  

First, 
cannabis use can acutely induce psychosis similar to the positive symptoms of SZ \cite{muller2008can, cortes2015psy}, 
and greatly increases the risk of SZ, with a 4-fold and 2-fold increase in heavy and average users, respectively \cite{marconi2016met}. 
Second, 
acute stress induces ECB synthesis, release and signaling, and chronic stress induces chronic ECB synthesis \cite{crosby2012int}.
Third, \sz\ is strongly associated with parental, prenatal, perinatal, and early life stress \cite{macdonald2009we, varese2012chi}. 
Finally, 
both acute and chronic stress can induce long-term epigenetic changes to the ECB system \cite{dAddario2013epi},
which can be heritable \cite{meccariello2020epi}.


These data provide a clear motivation for examining the hypothesis that stress-induced chronic ECB synthesis and/or signaling is the core cause of \sz. 

\smallest{PUFAs}
We will see right below that chronic ECBs can indeed explain the symptoms of SZ. However, before concluding that ECBs are the culprits in SZ, it should be asked whether the problem might be upstream of ECBs. 

The main ECB linked to SZ is N-arachidonoyl\-ethanolamine (AEA), whose synthesis requires the omega-6 (om6) PUFA arachidonic acid (ARA). A study of the literature quickly reveals a large body of evidence showing the involvement of ARA in SZ (see below). ARA is released from membranes by \pla 2 (PLA2), which also releases omega-3 (om3) PUFAs, mainly docosahexanoic acid (DHA), which are also known to be involved in SZ. Moreover, PLA2 is recruited by stress and \infl, and induces the synthesis of the other major ECB, 2-arachidonoylglycerol (2AG). 

\smallest{ECB, ARA and DHA mechanisms}
The main mechanistic effects of ECBs are well-known and supported by vast evidence (see Section~\ref{sec:ecbs}). 
First, ECBs disconnect the synthesizing cell from the neural network by suppressing presynaptic inputs (both GABAergic and glutamatergic).
Second, ECBs disconnect the cell from the \ecell\ environment by suppressing ion channels (calcium, potassium, cholinergic, serotonergic, others). 
Third, ECBs increase the cell's intrinsic \xtblty\ via elevated \icell\ \catwo, reduced GABAergic inputs, and oscillating currents. 
Finally, many of these effects are receptor-\indep, requiring only ECB synthesis. 
Importantly, like ECBs, both ARA and DHA (and its precursor eicosapentaenoic acid (EPA)) suppress external cellular inputs, and ARA increases \icell\ calcium and intrinsic \xtblty. Thus, 
\itemb
\item {ARA, ECBs and DHA decrease coordinated network activation, and ARA and ECBs increase spontaneous activation.} 
\iteme

In light of the shared effects of ECBs, ARA, and DHA, our initial hypothesis can be revised to assume that chronic stress-induced PUFAs are the cause of SZ. 


\smallest{Explaining the SZ symptoms}
Given these accepted mechanisms of action, can chronic PUFAs explain the symptoms of \sz? It turns out that they can do so in a straightforward manner. \thn\ explains SZ symptoms as follows.  
\itemb
\item
Positive symptoms occur due to \spont\ activation of neurons representing sensory experience (e.g., in the case of audition, in auditory and frontal cortices and \hippo), which then activate others to form sensory sequences (e.g., voiced sentences). Since cells are disconnected from the network, this activation is not restrained by neurons representing the spatial and temporal context or by neurons supporting executive function. Instead, it is guided by the probability of \spont\ activation, which is higher in the ECB-rich threat areas. This explains why psychosis is commonly accompanied by a feeling of threat and by paranoid delusions. 

Note that perception with a sense of reality in the absence of external sensory inputs is quite common. This is what happens when dreaming, 
and stimulation of the superior temporal gyrus readily induces auditory hallucinations \cite{scott2020int}. 

\item 
Negative symptoms occur because neurons that support volitional movement and social interaction are disconnected and are difficult to activate in a controlled manner. More specifically, ECBs and ARA directly oppose acetylcholine (ACh) signaling, which is crucial for movement. Opposition to ACh also explains eye pursuit abnormalities, a known marker of SZ. In addition, DA and norepinephrine (NEP) neurons are also disconnected and hyperexcitable, impairing vigorous volitional movement and arousal. 

\item
Cognitive symptoms occur because cognition depends on coordinated network activity. 
\item
Mood symptoms can occur prodromally simply because the person feels that their cognitive, motor and social performance is impaired. Such symptoms are naturally exacerbated following psychosis. However, they can also occur due to excessive \spont\ \xtblty\ of neurons in stress areas. Mood symptoms combined with hyperactive threat inputs and decreased cognitive control also explain occassional violence and suicidality. 
\iteme

Since the released PUFAs, especially ECBs, have a role in brain \plast, their chronic activation constitutes a dysfunction that impairs brain development and \plast. 


\smallest{I-PUFAs: the final hypothesis}
ARA is the substrate for prostaglandins (PGs), thromboxanes (Tbxs), and leukotrienes (Lkts), in addition to being a substrate for ECBs. DHA is a substrate for a family of derivatives known as specialized pro-resolving mediators (SPMs) (mainly lipoxins, resolvins, protectins) \cite{serhan2014pro}. 
We need to consider the involvement of these families in SZ. 

Examining the evidence, three data points stand out. First, SZ patients show a decreased niacin skin flush response, which is known to be induced by PGs. Second, many autoimmune diseases, including those linked to Lkts, are associated with increased risk of SZ, but rheumatoid arthritis (RA), which is strongly linked to PGs, is associated with decreased risk. Third, there is considerable evidence showing decreased PG synthesis in SZ. Based on these  and other data (see below), I hypothesize that PGs are not involved in SZ, while Lkts probably are. 

What is the factor that determines whether an agent has a role in SZ? A common distinction is that between pro-\infl\ and anti-\infl\ agents. However, both PGs and Lkts can show both effects, depending on context. I propose that the important distinction is the nature of stimulation by calcium, which coincides with the family's role in the stress response. 

I distinguish PUFAs and PUFA products into three types, those stimulated by \ecell\ \catwo\ influx (E-PUFAs), those stimulated by \icell\ \catwo\ (I-PUFAs), and those suppressing the other two (SPMs). E-PUFAs are associated with the initial acute stage of stress, and I-PUFAs are associated with prolonged stress and stress-induced \plast. 
SPMs terminate stress and its induced processes. 
I-PUFAs overlap to a large extent with the products stimulated by one particular enzyme, iPLA2. 

Most PGs are E-PUFAs, and ECBs, Lkts and DHA are I-PUFAs. ARA belongs to both groups. E-PUFAs promote strong neural activity, so do not signal like ECBs. The effects of ARA, DHA and Lkts are similar to those of ECBs. Since the \thn\ account of SZ symptoms depends on the suppression of network activity, it is valid for I-PUFAs but not for E-PUFAs. Thus, \thn\ posits the following:   

\itemb
\item
{\em SZ is caused by chronic upregulation of I-PUFAs, including ARA, DHA, ECBs, and LKts. The core cause is an upregulation of their synthesis by iPLA2, induced by prolonged stress during sensitive periods.}
\iteme

We will see below that ARA accumulation, rather than ECB signaling, may be the specific trigger of initial psychosis. 

I-PUFAs increase \icell\ \catwo, thereby stimulating 2AG (and AEA) synthesis. 
However, AEA decreases 2AG synthesis \cite{maccarrone2008ana}. These conflicting effects may explain why 2AG is not as strongly linked with SZ as AEA. 



\smallest{Evidence}
Is there evidence supporting this theory in addition to the stress- and cannabis-related data mentioned above? A study of the literature provides a definite affirmative answer (detailed in Section~\ref{sec:evidence}).  
\itemb
\item There is very strong lipid-based evidence, consistently reported over decades, for chronically increased release of ARA and DHA in SZ.

\item There is very strong evidence for chronic ECB synthesis and signaling in SZ.

\item There is clear evidence for reduced coordinated network activity and neural hyper\xtblty\ in SZ, affecting neuromodulators, hormones, and neurotransmitters. 
\item There is lipid- and immunity-based evidence that E-PUFA responses are blunted in SZ. 

\iteme

\smallest{Other triggers of psychosis}
Psychosis is a relatively common condition and can be induced by several agents, most sharing mechanisms of action with ECBs. 
DA (e.g., amphetamines) can induce psychosis by activating dopamine D2 receptors \cite{glasner2014meth}. 
DA induces AEA release, and its the brain distribution of D2 and its signaling cascade substantially overlap with that of the type 1 cannabinoid receptor (CB1), so the effect of activating D2 is partially similar to that of ECBs. 
Muscarinic ACh \antag s such as scopolamine induce psychosis by opposing ACh, which is also strongly opposed by ECBs. 
Salvinorin A is the primary psychoactive component of the hallucinogenic plant salvia divinorum. It is a selective agonist of the kappa opioid receptor, whose distribution and mechanism of action partially overlap those of CB1 \cite{walentiny2010kap}. 
NMDAR \antag s such as PCP and ketamine can induce psychosis \cite{hu2015glut}. NMDARs are the main receptors supporting prolonged \catwo-mediated execution, and their antagonism disconnects the network. 
Glucocorticoids (GCs) are routinely given during surgery to reduce post-surgery nausea and vomiting and to relieve pain, but can cause psychosis in almost 20\% of cases \cite{ross2012ste}. This happens because GCs recruit ECBs. 

Agonists of the serotonin receptor type 2a (SER2a) such as LSD are known psychedelics. SER2a cooperates with CB1 as well, but can facilitate psychosis via an additional route. It is present on thalamocortical inputs \cite{barre2016pre} (which also produce 2AG \cite{wu2010req}), and its strong activation can induce cortical activations similar to that induced by real sensory inputs. In addition, SER2a excites pyramidal neurons, mainly in prefrontal cortex (PFC) \cite{jakab2000SER}, and increases \icell\ \catwo, which can recruit ECBs. 

See below for psychosis induction in \epil\ and by NSAIDs and anti-malaria drugs. 

\smallest{\Antip s}
\thn\ explains the mechanisms and deficiencies of existing SZ medication as follows. All current \antip s are D2 \antag s, and atypical \antip s are also SER2a \antag s. 
 D2 and SER2a can promote psychosis via the mechanisms described above. 
 Antagonizing D2 suppresses its induction of AEA release, while antagonizing SER2a reduces the drive for sensory hallucinations by suppressing thalamocortical inputs. 
However, these drugs are not effective against other symptoms, because I-PUFAs affect many neurons not expressing D2 or SER2a. D2 antagonism can also lead to \desensit\ of the ACh system to induce tardive dyskinesia, and both types of \antag s may exacerbate chronic I-PUFAs in the longer term by increasing PUFA release. 

\smallest{Summary}
Stress normally recruits PUFAs and their products. Prolonged stress recruits I-PUFAs. In some cases, e.g., when prolonged stress occurs at sensitive periods, this recruitment can become chronic. Chronic I-PUFAs can explain SZ symptoms and pathophysiological data. The stress-induced alterations include epigenetic changes that are heritable, explaining the familial association data of SZ. 

\refsection{PUFAs and PUFA Derivatives in Health}{ecbs}

\smallest{General}
The major membrane lipids are glycerophospholipids, sphingolipids, and sterols. In mammals, the most abundant glycerophospholipids are phosphatidylcholine (PC), phosphatidyl\-ethanolamine (PE) and phosphatidylinositol (PI).
PC and PE contain a choline or ethanolamine head group (respectively) and fatty acids, usually one saturated 
and one unsaturated. 
The latter is commonly a polyunsaturated fatty acid (PUFA), with both omega 6 (om6) and omega 3 (om3) PUFA being common \cite{bazinet2014PUFA}. 
The most abundant PUFAs are ARA (om6) and DHA (om3), both located at the sn-2 position of phospholipids. 
ARA and DHA are produced by elongation and desaturation from linoleic acid (LA) and alpha-LA (ALA) (respectively), which must be procured from the diet. 
PEs usually contain more ARA and DHA than PCs. 

\smallest{PLA2}
The PLA2 family is comprised of calcium-dependent (cPLA2), secreted (sPLA2), calcium-\indep\ (iPLA2), and other less major members \cite{wilkins2008gro}. 
All PLA2s hydrolyze the fatty acid at the sn-2 position of phospholipids (mainly ARA and DHA), releasing free fatty acids and lysophospholipids. 
iPLA2 induces both ARA and DHA release, while cPLA2 mainly releases ARA for PG synthesis \cite{wilkins2008gro}. 
While iPLA2 activity is mainly associated with PE, it also hydrolizes PC \cite{barbour1999reg}
and PI \cite{negre1996cha}. 

The activation of cPLA2 and sPLA2 is triggered by \ecell\ \catwo, while reduced \er\ \catwo\ (associated with increased \icell\ \catwo\ or with calcium depletion) activates iPLA2 \cite{rosa2009int}. 
Thus, iPLA2 is an I-PUFA enzyme. 
GCs stimulate cPLA2 expression when signaling via cAMP (i.e., at the initial stage of GC signaling), and downregulate it otherwise \cite{guo2010ind}, freeing ARA to be used by iPLA2. 
iPLA2 transcription is stimulated by sterol regulatory element-binding proteins (SREBPs) \cite{ramanadham2015cal}. 

\smallest{ECBs}
The ECB system comprises a family of signaling lipids, their synthesis and degradation enzymes, and receptors \cite{katona2012mul}. 
Most research focuses on two ECBs, 
AEA and 2AG, and on the brain's CB1 receptor. CB2 is mainly expressed by immune cells. 

The synthesis of AEA and 2AG is stimulated by \icell\ \catwo\  \cite{katona2012mul} and by factors that increase it, including stress (see below), and requires ARA. PE and ARA form N-arachidonoyl phosphatidylethanolamine (NAPE), which serves as a substrate for direct AEA synthesis via NAPE phospholipase D (NAPE-PLD), or indirect synthesis via PLA2 \cite{sun2004bio} 
or phospholipase C (PLC) \cite{liu2006bio}. 
AEA is hydrolized by fatty acid amide hydrolase (FAAH). Importantly, FAAH also induces the reverse reaction, producing AEA from ARA and ethanolamine at larger substrate concentrations \cite{tsuboi2018end}.
2AG contains ARA and glycerol, and is formed from diacylglycerol (DAG) by diacylglycerol lipase (DAGL). 
PLC mediates the production of DAG from phosphatidylinositol 4,5-bisphosphate (PIP2). 2AG is hydrolized by monoacylglycerol lipase (MAGL). 

%
CB1 is mainly located in presynaptic axons, although it also exhibits postsynaptic and \mitoch\ localization\cite{busquets2018cb1}. 
In axons, it is mainly expressed by GABAergic inhibitory interneurons (IINs). However, ECBs directly affect pyramidal neurons as well \cite{maroso2016can}. 

The brain areas involved in stress and memory, including the \amg, \hippo, and PFC, are especially enriched in ECBs. 

\smallest{Other ARA products}
As mentioned above, ARA is a substrate for ECBs, PGs, Tbxs, and Lkts. PGs and Tbxs are produced via cyclooxygenases (COX), mainly using cPLA2-released ARA, and most of them promote acute \infl. 
The key PG enzymes COX1 and COX2 can also produce PG from AEA and 2AG. 
ECB and PG synthesis seem to compete for ARA, with GCs pushing ARA utilization towards AEA \cite{goppelt1989glu}, most likely via destabilization of COX2 mRNA \cite{ristimaki1996dow}. 
Non-steroidal anti-inflammatory drugs (NSAIDs) such as aspirin and ibuprofen suppress COX2, thereby increasing the utilization of ARA by iPLA2. 
PGE2 opposes CB1 \cite{xia2012pro}, while 2AG opposes COX2 expression \cite{zhang2008end}. 
Thus, the inflammatory PGs and Tbxs are E-PUFAs. 

Lkts are produced from ARA by lipoxygenases (LOX, mainly LOX5), and can both promote and oppose \infl. 
Lkt synthesis is usually described as being stimulated by cPLA2. However, Lkt production requires \catwo\ influx via calcium release activated channels \cite{chang2004clo, finney2009leu}, whose activation in turn requires iPLA2 \cite{smani2003ca2}. Thus, Lkts are I-PUFAs. 
LOX5 is also activated by substrate availability \cite{brash2001ara}. 

\smallest{DHA}
The om3 PUFA DHA is used by lipoxygenases (LOX, mainly LOX15 but also LOX12) to produce a family of derivatives known as specialized pro-resolving mediators (SPMs) (e.g., lipoxins, resolvins, protectins) \cite{serhan2014pro}, which actively terminate \infl\ and stress. Most DHA is used for beta-oxidation or reesterified into membranes \cite{bazinet2014PUFA}. 

\smallest{Mechanistic effect}
Most aspects of ECB signaling are well-known. ECBs are mainly retrograde messengers, acting on receptors located presynaptically to the releasing cell. They can also act on the cell itself in an autocrine manner. They have both receptor-dependent and receptor-\indep\ effects. 

CB1 is a G protein-coupled receptor (GPCR) mainly bound to Gi/o. When activated, it inhibits adenylate cyclase to decrease cAMP and neuro\transm\ release, and increases G protein-coupled inwardly rectifying \kplus\ (GIRK) channels \cite{busquets2018cb1}. 
Since CB1 is mainly expressed on presynaptic axons, its activation serves to disconnect neurons from their inputs. Since it is mainly expressed on IIN axons, its net effect is to reduce external inhibition. 

ECBs also have direct receptor-\indep\ effects. They directly suppress ion channels, including calcium, nicotinic ACh, serotonin, and some potassium channels \cite{oz2006rec}. 
AEA and a related molecule bind muscarinic ACh receptors, and inhibit agonist binding to it at high doses (with facilitation at low doses) \cite{lagalwar1999ana}. 
ECBs also act on DA, NEP, ACh and SER neurons (see under evidence below). 


The transient receptor potential vanilloid 1 (TRPV1) channel is a general cation channel mainly expressed in sensory input paths (e.g., dorsal root ganglia). AEA is a TRPV1 ligand that induces rapid channel \desensit, resulting in a relatively small increase of \icell\ \catwo\ and autocrine cellular excitation \cite{varga2014ana}. 
%
%
ECBs can also utilize the GPR55 receptor to increase \icell\ \catwo\ \cite{sharir2010pha},  
and increase intrinsic postsynaptic \xtblty\ via Ih currents \cite{maroso2016can}. 


Both ARA and DHA inhibit voltage-gated calcium and sodium channels \cite{boland2008pol, vreugdenhil1996pol}, 
nicotinic ACh receptors \cite{bouzat1993eff}, 
and GABA currents (DHA at chronic doses) \cite{nabekura1998fun}. 
Both trigger \catwo\ release from \icell\ stores \cite{meves2008ara, rosa2009int}, 
and DHA directly activates TRPV1 and other TRP channels \cite{matta2007trp}. 

LktC4 increases the potassium M-currents in \hippo\ pyramid neurons, thereby hyperpolarizing them and directly opposing muscarinic ACh signaling \cite{schweitzer1990ara}. 
Both LktC4 \cite{wang2021bas} and LktB4 \cite{fernandes2013sup} promote itch via neural circuits, including TRPV1 \cite{fernandes2013sup}. 


SPMs reduce brain activity during strong stress \cite{bazan2011end} 
and increased ECB levels in a pain model \cite{khasabova2020int}, 
but are different from other I-PUFAs in that they reduce intrinsic \xtblty\ by suppressing TRP channels, including TRPV1 \cite{payrits2020res}. 

In summary, 
\itemb
\item
{ I-PUFAs and SPMs disconnect cells from network inputs and from the extracellular environment, and I-PUFAs increase intrinsic neural \xtblty}. 
%
\iteme

\smallest{Function}
The possible roles of ECBs have been widely discussed, with the leading candidates being \plast,
neuroprotection,
and energy \homeos\ \cite{cristino2014ECB}. 
For the purpose of this paper, we only need their established role in synaptic \plast. \Plast\ processes involve massive DNA access, which is dangerous with \ecell\ \catwo\ influx and the activation of various signaling cascades triggered by inputs from other neurons. ECBs disconnect cells from the environment in order to minimize this risk, and promote \icell\ \catwo\ (and thus \spont\ \xtblty) to streamline ATP production and protein synthesis activities in \mitoch\ and the \er. 


Other I-PUFAs are also involved in stress-induced \plast. Inhibition of iPLA2 prevents \plast, which can be rescued by ARA or DHA \cite{fujita2001doc}. 
LOX5 knockout mice exhibit motor deficits associated with abnormal cortical and hippocampal synapses \cite{barbosa2022mic}. 
The DHA metabolite synaptamide stimulates neurite growth and synaptogenesis \cite{kim2011syn}. 
SPMs were investigated mainly in the context of \infl\ resolution, but there is some evidence that they promote brain \plast\ \cite{wang2015res, shalini2018dis}. 


The functional difference between AEA and 2AG has been widely discussed, but for our purposes it suffices to note that 2AG release is rapid while AEA release is slower and of longer duration. Thus, it is natural to assume that 2AG and AEA support the initial and later phases of \plast, respectively. 
Indeed, the triggering phase of stress involves suppression of FAAH and reduction of \amg\ AEA \cite{gray2015CRH}. 



\refsection{PUFAs and PUFA Derivatives and Stress}{stress}

The interaction between stress and the ECB system is very well-known, and is summarized here. The term `stress' describes the states induced by threatening inputs or internal danger to cells (e.g., due to radiation, energy deficiency, calcium overload, immunity-inducing pathogens). 
These are the major states that trigger \plast\ processes, whose role is to ensure that the organism is better prepared to execute a response the next time that the threat or danger occur. 

Threat inputs trigger two types of stress responses, early and late. The early response, mediated by NEP, recruits energy and neural resources to support quick and vigorous execution. The late response, mediated by \gx s (GCs, cortisol in humans), is meant to occur after execution termination, and its goal is to implement the \plast\ changes that follow execution. In both cases, the appropriate agents (NEP and GC) are recruited by corticotropin-releasing hormone (CRH). CRH represses AEA (as noted above), and GC recruits it (see below). 

Internal cellular stress recruits and signals via oxidative species, mainly superoxide and hydrogen peroxide (H2O2). 

The relevant facts for SZ are: 
 
\itemb
\item {\em Stress recruits ECBs} \cite{gray2014ECB, balsevich2017ECB}. 
GCs can increase \icell\ calcium \cite{chameau2007glu} and directly stimulate PLA2 expression \cite{guo2010ind}. 
As noted above, they shift ARA utilization from PGs to ECBs. 
H2O2 increases cytosolic calcium \cite{kuster2010redox}, upregulates CB1 expression, and downregulates FAAH expression \cite{wei2009pre}. 


\item {\em Stress can have long-term effects, mainly when it occurs in sensitive developmental periods} such as pregnancy, perinatal, early life, and \adol\ \cite{sandman2016neu, meyer2019neu, babenko2015str, short2019ear, eiland2013str}. The induced effects include increased risk for many disorders, and alterations in the stress system itself. 


\item {\em The long-term effects of stress can involve \epig\ alterations and be heritable} \cite{szyf2015non, dick2021str}. 


\item {\em Stress can have long-term effects on the ECB system} \cite{crosby2012int, goldstein2021ear, rusconi2020end, demaili2023epi}. In this sense, exposure to cannabis is a stress state \cite{zumbrun2015epi}. 

\item {\em The ECB system can be affected by \epig s}, including after stress \cite{franklin2010epi, xia2012pro, dAddario2013epi, hong2015epi, wiedmann2022dna}. 

\item {\em Stress and cannabis effects on ECBs can be heritable} \cite{franklin2010epi, meccariello2020epi}. 
\iteme

The major long-term change induced by stress is increased \xtblty\ of circuits involved in stress responses, which may seem to imply an upregulation of all stress system components. However, the CRH-GC axis suppresses itself via negative feedback, and chronic activation of GC and ECB receptors can desensitize them \cite{hapgood2016glu, hendrik2020con}. As a result, chronic stress usually results in increased agent levels (GC, ECBs), but can induce both increased and decreased receptor density and signaling. 
Indeed, stress has been reported to upregulate all of the components of the ECB system \cite{robinson2010eff, sciolino2010soc, zamberletti2012chr, marco2014con, schneider2016adv, boero2018imp, demaili2023epi}, 
but also to downregulate CB1 expression \cite{franklin2010epi, sciolino2010soc, zamberletti2012chr, crosby2012int}. 
 

SPMs have not been investigated to the same extent as ECBs, but seem to show similar properties. SPMs are recruited by acute \infl\ (immune stress) and by GCs (external stress) \cite{uz2001glu}, their alteration during pregnancy affects the newborn \cite{elliott2017rol}, and their paths are affected by \epig\ mechanisms \cite{imbesi2009lox}. 

\refsection{Evidence for \thn }{evidence}
In addition to the cannabis and stress evidence summarized above, \thn\ is supported by large bodies of evidence with respect to lipids, immunity, pain, the ECB system, neuromodulators, hormones,  neurotransmitters, and cortical activity in SZ. 



\subsection{PUFAs}

There is overwhelming evidence showing lipid dysregulation in SZ. Here I show that the vast majority of this evidence points to increased liberation of ARA and DHA in SZ. 
The main evidence is presented below\footnote{A large number of papers reported lipid-related measurements in SZ, so it can be assumed that for each change listed here there are several papers that have not detected it. Explicitly opposite results and some of the null results are reported here.}.  

\smallest{PLA2} 
Increased PLA2 activity was directly measured and reported in many papers \cite{gattaz1987inc, noponen1993ele, gattaz1995inc, ross1997inc, ross1999dif, tavares2003inc, smesny2005inc, smesny2010pho, smesny2011pho, xu2019inv, yang2021dys, talib2021inc, li2022imp}. Increased protein \cite{xu2019inv} and mRNA \cite{yang2021dys} were reported as well. Measurements were in plasma or serum \cite{gattaz1987inc, noponen1993ele, ross1997inc, xu2019inv, yang2021dys}, platelets \cite{gattaz1995inc, talib2021inc}, red blood cells \cite{li2022imp}, or brain \cite{ross1999dif}. The results pertained to SZ but not other psychiatric disorders \cite{gattaz1995inc, ross1999dif} or occurred in several disorders \cite{noponen1993ele}. Results were not due to medication \cite{gattaz1987inc, ross1999dif, tavares2003inc}, in some cases only occurring in first-episode psychosis (FEP), medicated or not \cite{smesny2011pho, smesny2010pho, smesny2005inc}. They were also reported for people at ultra-high risk \cite{talib2021inc}. 
Most papers reported increased iPLA2 \cite{ross1997inc, ross1999dif, smesny2005inc, smesny2010pho, smesny2011pho, xu2019inv, talib2021inc}. Two of these reported also increased \cite{talib2021inc} or decreased \cite{ross1999dif} cPLA2. One paper found higher plasma cPLA2 with no clear iPLA2 changes \cite{yang2021dys}. 
One paper reported normal activity in drug-naive patients \cite{albers1993pho}. 

Two additional facts support the involvement of iPLA2 in psychosis. 
First, the prevalence of psychosis in \epil, especially temporal lobe \epil, is much higher than in the general population \cite{clancy2014pre}. Indeed, there is increased \hippo\ iPLA2 (but not sPLA2 or cPLA2) in patients with vs.\ without psychosis \cite{gattaz2011inc}. In many cases, \epil\ psychosis involves an ongoing small seizure activity \cite{elliott2009del}, which would increase non-acute \icell\ \catwo. 
Second, NSAIDs, which suppress COX2 and thereby increase the utilization of ARA by iPLA2, can induce psychosis \cite{onder2004nsa}.

A little-known epidemiological fact supports a role for higher PLA2 activity in SZ. Anti-malaria drugs such as chloroquine and quinacrine are non-selective PLA2 inhibitors \cite{lu2001dif}, and 
some of the lowest mean SZ prevalence rates are in sub-Saharan Africa, where anti-malaria drugs are highly used  \cite{charlson2018glo}\footnote{However, similarly low rates were reported for North Africa and the Middle East.}.

\smallest{ARA}
Decreased membrane ARA was reported in many papers \cite{peet1995dep, yao1996abn, yao2000mem, khan2002red, arvindakshan2003ess, reddy2004red, kemperman2006low, mcnamara2007abn, sethom2010pol, rice2015ery, hamazaki2016fat, wang2018alt, alqarni2020com, zhou2020red}. Measurements were in blood \cite{peet1995dep, yao1996abn, khan2002red, arvindakshan2003ess, reddy2004red, kemperman2006low, sethom2010pol, rice2015ery, wang2018alt, alqarni2020com, zhou2020red, li2022red} and brain \cite{yao2000mem, mcnamara2007abn, hamazaki2016fat}. Reports included FEP or drug-free patients \cite{khan2002red, reddy2004red,mcnamara2007abn, zhou2020red, li2022red} and people at ultra-high risk \cite{rice2015ery, alqarni2020com}. 
This result was also reported in meta-reviews \cite{hoen2013red}.
Significantly increased ARA \metab\ and turnover were reported in two papers \cite{yu2022nia, das1998inc}. 
Plasma showed higher cPLA2 activity, with unchanged ARA membrane incorporation enzyme \cite{yang2021dys}. 
Moreover, red blood cell ARA and LA were inversely correlated with symptom severity \cite{berger2019rel}. 
These results point to increased ARA release and utilization in SZ. 

One paper reported CSF ARA, and as expected, it was negatively correlated with symptoms \cite{kale2008opp}. This paper reported no ARA differences in red blood cells. 
Normal ARA in skin fibroblasts was also reported \cite{mahadik1996pla}. 

Increased levels of red blood cell LA (the ARA precursor) were associated with increased conversion of people at ultra-high risk \cite{frajerman2023mem}. 
Red blood cell ARA was positively correlated with total \icell\ PLA2 activity in people at ultra-high risk who converted to psychosis, and negatively in non-converters \cite{smesny2014ome}. 
Both patients and their siblings show increased red blood cell ARA, DHA and DPA \cite{medema2016lev}. 
Increased ARA vs.\ om3 PUFA was reported to correlate with hostility symptoms \cite{watari2010hos, szeszko2021lon}. 
These results indicate that increased availability of ARA increases SZ risk.  


\smallest{Prostaglandin} 
PG data in SZ were reported in several papers, many reporting decreased levels. In the brain, significantly decreased frontal cortex PGE synthase was reported, with no COX changes \cite{maida2006cyt}. However, this was shown by depression and bipolar patients as well. 
Dorsolateral PFC COX1 was decreased (41\%) in 
SZ brain but not in depression  \cite{dean2018stu}. 
Blood antibodies from patients trigger PGE2 production and induce COX1 mRNA in rat frontal cortex \cite{ganzinelli2010aut}. 

In blood, PGE1 synthesis in platelets can be stimulated in controls and patients with depression, but not in SZ patients, especially acute ones \cite{abdulla1975eff, kaiya1991pro}. 
Decreased COX2 mRNA was reported, but only in patients older than 40 years old \cite{tang2012dif}. 
Plasma COX2 and PGE2 were reported higher in chronic medicated patients vs.\ controls and bipolar patients, while COX1 was higher in both SZ and bipolar \cite{garcia2018reg}. 
Serum COX-produced PGs were decreased in FEP \cite{wang2018alt}. 
Serum PGE2, 15d-PGJ, and \infl\ marker CRP were reported significantly lower \cite{yuksel2019ser}. 
Serum COX- produced PGs and LOX12 metabolites were reported lower, while EPA and LOX15 metabolites were higher in pre-medication SZ \cite{wang2018alt}. 
Plasma COX2 and TbxA synthase were reported reduced in SZ, while COX1 (and cPLA2) mRNA was higher \cite{yang2021dys}. 


One paper reported unchanged COX2 mRNA in frontal cortex and blood mononuclear cells \cite{fromer2016gen}. 

Several papers reported increased PGs, all in blood. 
Patients had increased plasma PGE2, with patients with higher levels having more hallucinations \cite{kaiya1989ele}. 
The higher ARA \metab\ in FEP platelets was attributed to PG \cite{das1998inc} (but this has not been directly shown). 
SZ patients had higher plasma PGE2 and decreased 15d-PGJ2 and its nuclear receptor PPARg in red blood cells \cite{martinez2011ant}. 
Higher COX1 and PG3 receptor mRNA was reported along with the decreased COX2 mRNA in patients older than 40 years old cited above \cite{tang2012dif}. 
FEP blood mononuclear cells showed increased COX2 (PGE2 only trend), and decreased 15d-PGJ2 \cite{garcia2014pro}. 
PGE2 and COX2 were increased with (very large variability of PGE2), and PGJ2 decreased, in FEP blood \cite{cabrera2016cog}. 
Plasma PGE2 and thromboxane B2 were higher in ultra-high risk, and correlate with negative and positive symptoms respectively \cite{pereira2022cox}. 


Niacin (also known as nicotinic acid and vitamin B3) normally induces PG and a skin flush response \cite{nadalin2010nia}. This response has been consistently shown to be reduced in SZ \cite{nadalin2010nia, yang2021dys}, indicating either reduced PGs or desensitized receptors (PG receptors are GPCRs). 
In one paper, the decrease was associated with increased PLA2 activity \cite{tavares2003inc}. 


As mentioned above, GCs (and thus prolonged stress) push ARA utilization towards AEA rather than PGE2. This explains the generally decreased PGs in SZ. Nonetheless, precursor availability by itself most likely stimulates PG synthesis \cite{dong2016dif}, explaining the reports of increased PGs in SZ.  
It is also possible that both paths are enhanced, with the I-PUFA more strongly so. 

\smallest {DHA, om3}
Brain membrane DHA levels were reported unchanged in 9 papers \cite{lacombe2018bra} and decreased in one \cite{mcnamara2007abn}. 
Several papers reported decreased blood DHA, in drug-native FEP (including a meta-review) \cite{khan2002red, arvindakshan2003ess, reddy2004red, sethom2010pol, mcnamara2013adu, mcevoy2013lip, hoen2013red, wang2018alt}, ultra-high risk \cite{alqarni2020com}, and in general \cite{peet1995dep, assies2001sig, khan2002red, kemperman2006low, kale2008opp, wood2015dys, li2022red}. 
Significantly decreased DHA, total om3 PUFAs, and lower utilization of PEA into DHA were reported in FEP skin fibroblasts \cite{mahadik1996pla, mahadik1996uti}. 

Decreased DHA, and a high om6/om3 ratio, are associated with symptoms \cite{watari2010hos, berger2017om, berger2019rel, szeszko2021lon, li2022red}. 
Treatment that improved symptoms also increased the reduced om3 and om6 levels \cite{li2022red}. However, the om3 index was decreased after treatment.  
One paper reported increased red blood cell DHA and ARA in patients and their siblings \cite{medema2016lev}. 
One paper reported no om3 changes in ultra-high risk \cite{rice2015ery}. 

These data indicate higher release and utilization of DHA in blood, and a protective association of higher om3 or om3/om6 ratio, in blood but not brain. This might show that the underlying PUFA problem is manifested more strongly in blood, and that symptoms are directly related to ARA but not to DHA. 

\smallest{ARA precursors and metabolites}
Membrane PE and PC (mainly PE) are widely reported to be decreased in SZ \cite{pettegrew1991alt, keshavan1993ery, mahadik1994pla, stanley1995viv, yao2000mem, schmitt2001eff, ponizovsky2001pho, ryazantseva2002cha, schmitt2004alt, kaddurah2012imp, mcevoy2013lip, wang2019met, alqarni2020com, wang2021cha}, 
and phosphodiesters (ARA metabolites) are reported to be increased \cite{pettegrew1991alt, stanley1995viv, schmitt2001eff, wang2019met, wang2021cha, li2022imp, song2023pot}. 
These results were reported in brain (dorsolateral PFC, caudate, thalamus) \cite{pettegrew1991alt, stanley1995viv, yao2000mem, schmitt2004alt}, blood \cite{keshavan1993ery, schmitt2001eff, ponizovsky2001pho, ryazantseva2002cha, kaddurah2012imp, mcevoy2013lip, wang2019met, alqarni2020com, wang2021cha, li2022imp, song2023pot}, 
and skin fibroblasts \cite{mahadik1994pla}, 
including in early and drug-free stages \cite{stanley1995viv, kaddurah2012imp, mcevoy2013lip, song2023pot} and ultra-high risk \cite{alqarni2020com}. 
One group reported decreased serum lysoPC \cite{wang2019met, wang2021cha}. 
One paper reported significant depletion of om3 and om6 PUFAs from internal PE (recall that iPLA2 activity is internal), with increased PE in the external membrane leaflet of red bloow cells \cite{nuss2009abn}. 

These results strongly support increased ARA release in patient brain and blood cells, including in prodromal and early drug-free stages. 

\smallest{Inositol}
Increased PI turnover was reported in patient brain \cite{atack1998cer, shimon1998ino, jope1998sel} (but also in depression \cite{atack1998cer}), blood \cite{essali1990pla, das1992ino, yao1992inc, yao1996abn,  rybakowski1997inc}, and skin fibroblasts \cite{mahadik1994pla}. 
PI enzymes were altered in postmortem brain in SZ \cite{kunii2021evi}. 
This may indicate that 2AG is also hyperactive in SZ, not just AEA. Indeed, there is an iPLA2 variant that hydrolizes PI \cite{negre1996cha}. 

\smallest{LOX products (Lkts, SPMs)}
There is almost no data about LOX products in SZ. With respect to Lkts, 
One paper reported decreased serum LOX5 products \cite{wang2018alt}, and another reported reduced LOX5 activating protein in SZ vs.\ neural \degen\ disorders \cite{durrenberger2015com}. 
With respect to SPMs, decreased LOX12 products but increased LOX15 products \cite{wang2018alt}, 
increased LOX12 products \cite{yao1996abn}, 
downregulated LOX15b in three repositories \cite{piras2019per}, 
and an association with LOX12 mutations \cite{kim2010ass} were reported. 
No conclusions can be drawn from these results. 
In a mouse model, chronic social defeat stress increased basal ganglia LOX12 metabolites after one week (but not after a day), significantly more in resilient mice \cite{akiyama2022chr}. 
This is not a validated model for SZ, but it does show that prolonged social stress recruits SPMs. 


\smallest{Genetic data}
Genetic associations of iPLA2 with SZ have been reported, all of them relatively weak \cite{junqueira2004all, yu2005gen, tao2005cyt}. 
A combined iPLA2-cPLA2 heterozygosity was associated with a 2x higher risk \cite{nadalin2017ass}. 
Five SREBP1 and two SREBP2 genetic markers were associated with SZ in German, Danish, and Norwegian samples of total size of about 1500 people \cite{leHellard2010pol} (recall that SREBPs, mainly SREBP1, stimulate iPLA2 transcription \cite{ramanadham2015cal}). 
Seven positive and five negative results were reported for cPLA2 \cite{tao2005cyt, barbosa2007ass, nadalin2008ban}. 

It should be noted that weak genetic associations of this kind have been reported for a large number of other genes, so this data does not provide strong support for the theory. 

\smallest{Obesity, diet}
SZ patients tend to be more obese than controls \cite{allison2009obe}. This is widely attributed to the effect of \antip s, but is commonly (not always) reported also in FEP \cite{allison2009obe, garrido2022pre}. 
A review found that patients exhibit higher caloric intake, without preference to any particular category \cite{strassnig2005diereview}. 
These data support \thn, since ECBs promote food intake and obesity \cite{cristino2014ECB}. 


Obesity can also be attributed to patient diet,  
as the natural diet of people with SZ seems to include a high consumption of fats \cite{strassnig2005die, dipasquale2013die}. 
However, the situation here is more complex.
Mammals cannot synthesize most PUFAs so must obtain them from the diet, mainly linoleic acid (LA) as the om6 precursor and alpha-LA (ALA) as the om3 one. The synthesis of om3 and om6 precursor-derived fatty acids competes for the same enzymes, so increasing intake of one of them can reduce the other to change the om6/om3 ratio \cite{bazinet2014PUFA}. 
The common view is that increased om6 vs.\ om3 PUFA increases risk.  
People at ultra-high risk who convert to psychosis, and FEP patients, showed a higher intake of om6 PUFA than controls \cite{pawelczyk2016ass, pawelczyk2017dif}, 
and significantly lower om3 intake in patients has been found \cite{kim2017low}. 
Normal red blood cell LA in FEP drug-naive patients, coupled with lower ARA, were reported in 5/6 papers \cite{mcnamara2009mod}, which indicates higher LA (om6) intake. 
In addition, in several clinical trials, om3 supplementation was modestly beneficial during the initial stages of the disorder \cite{bozzatello2019pol} or in patients with a low om3 baseline \cite{hsu2020ben}. 
These data generally support \thn , because ARA products (mainly ECBs) have a stronger I-PUFA effect on the brain than DHA products. Indeed, om3 supplementation reduces total \icell\ PLA2 activity \cite{smesny2014ome}, 
and decreases tissue levels of AEA and 2AG \cite{kim2013fat}. 
Thus, high om6 and/or low om3 intake (e.g., due to economic conditions) can increase risk. 

However, both om3 and om6 intake reduced risk in a large Swedish cohort \cite{hedelin2010die}, 
and 
higher om3 intake in FEP patients vs.\ controls was also found \cite{pawelczyk2017dif}. 
In addition, a negative correlation between SZ incidence and average om6 and ARA intake was found in 24 countries \cite{gao2023ass}. 

\thn\ explains these data by noting that intake patterns can also be compensatory. SZ involves excessive release of ARA and DHA from membranes, and this can promote higher PUFA intake (especially om6) to replace them. Ethanolamine, which is used by FAAH to produce AEA when ARA is excessive, must be obtained from the diet and is 3x more abundant in meat than fish \cite{fogerty1991com}. 
Hence, meat consumption opposes its depletion. 
In this view, patients consume higher amounts of fat because their cells need fat. The risk to the organism posed by complete collapse of cell membranes due to lack of PUFA is higher than the risk resulting from the symptoms of SZ. 


\smallest {Summary}
PLA2 (usually iPLA2) activity is definitely increased in SZ, accompanied by increased freeing of ARA and DHA. This activity is not due to medication. The freed ARA is most likely used for AEA and not PG (a conclusion strengthened by the immunity and pain data below). 
All this indicates that increased I-PUFA synthesis capacity is a core property of the disorder. 
The fact that the phenomenon occurs in both blood and brain supports an immune and/or \epig\ origin. 
Patients try to compensate for reduced membrane PUFA via their diet. 

\subsection{Immunity and pain}
Meta-reviews show evidence for immune involvement in SZ, including activation of brain microglia \cite{trepanier2016pos}, increased levels of CSF and blood \infl\ cytokines in FEP \cite{miller2011met}, and plasma biomarkers possibly predicting transition to psychosis in high risk \cite{khoury2018inf}. 

More specific information can be derived from immunity-related diseases. 
SZ is positively associated with psoriasis, atopic dermatitis, asthma, lupus, and some other diseases \cite{cullen2019ass, wang2018aut, chen2009prev, pedersen2012sch, tiosano2017sch}. 
There is a strong negative association with rheumatoid arthritis (RA) \cite{oken1999iss, cullen2019ass} 
(although a recent positive association in an all-Taiwan cohort was also reported \cite{wang2018aut}). 
It has long ago been noted that patients are less likely to have allergies and show decreased responses to histamine injections \cite{matthysse1975bio, heleniak1999his}. 

Importantly, PGs are known to promote RA \cite{mccoy2002rol} and allergies \cite{pettipher2008rol}, 
while Lkts promote asthma, psoriasis, and atopic dermatitis \cite{hallstrand2010up, ruzicka1986ski}. 
This implies reduced PGs in SZ, which supports the distinction made in \thn\ between E-PUFAs and I-PUFAs, and the \thn\ hypothesis that the excessive ARA in SZ is used for ECB and Lkt, rather than PG, production. 


SZ patients show decreased pain sensitivity \cite{horrobin1977sch, heleniak1999his, engels2014cli}. 
This can be explained in two ways, both supporting \thn . First, the TRPV1 channel is an important conveyor of pain, and is desensitized by I-PUFAs. Second, PGE2 is a major pain-inducing agent, and decreased pain accords with it being reduced in SZ. 


\subsection{ECB system}
All of the components of the ECB system are dysregulated in SZ, with the evidence pointing to a chronic increase. 

\smallest{ECB levels} 
Plasma and CSF ECB levels (mainly AEA but also 2AG) were reported to be dramatically increased in many papers \cite{leweke1999ele, deMarchi2003end, giuffrida2004cer, leweke2007ana, potvin2008end, koethe2009ana, muguruza2013qua, wang2018alt, koethe2019fam, potvin2020per, ibarra2022can, parksepp2022exp, romer2023bio}, 
including a systematic review and meta-analysis \cite{minichino2019mea}. 
These reports include FEP, drug-naive \cite{giuffrida2004cer, leweke2007ana}, prodromal \cite{koethe2009ana}, and drug-free \cite{muguruza2013qua, wang2018alt} patients, and patients evaluated in psychiatric emergency \cite{potvin2020per}. Cannabis use has been accounted for in some papers. 
Increased blood AEA was also reported in people at high-risk with childhood trauma (vs.\ high-risk without  trauma), with significant correlation of NAPE with symptoms \cite{appiah2020chi}. 

Lower AEA in some brain areas was reported too, with the same paper reporting higher 2AG \cite{muguruza2013qua}. 
Decreased serum 2AG was reported in a paper reporting increased N-acylethanolamines (which are ECB-like) \cite{parksepp2022exp}. 
Lower blood AEA and 2AG was reported in ultra-high risks vs.\ controls \cite{joaquim2022pla}. This is the only result in which both AEA and 2AG were reduced. Note that it is in blood, not CSF. 


Ethanolamine is markedly decreased in patient plasma and platelets \cite{wood2015dys}. This indicates higher AEA synthesis, which is stimulated by abudance of ARA and ethanolamine via the FAAH reverse reaction \cite{tsuboi2018end}. 

Thus, AEA (and 2AG) levels are increased in SZ, which supports \thn . Further support is provided by papers from Leweke and colleagues reporting dramatically higher CSF (but not serum) AEA in disorder states, but with higher AEA correlating with better, not worse, prognosis. The papers compared FEP drug-naive  patients vs.\ controls \cite{giuffrida2004cer}, FEP drug-naive patients with low frequency of cannabis use vs.\ high frequency use or controls \cite{leweke2007ana}, initial prodromal state vs.\ controls \cite{koethe2009ana}, and discordant twins vs.\ healthy twins \cite{koethe2019fam}. 
In two papers, AEA negatively correlated with psychotic symptoms \cite{giuffrida2004cer, leweke2007ana}, and in the other two, higher AEA was associated with reduced conversion to psychosis \cite{koethe2009ana, koethe2019fam}. 
The authors interpret their results as AEA being protective in SZ. In their view, AEA is high in risk states as a protective mechanism, and risk may transform to psychosis when it is not high enough \cite{leweke2017put}. 

The \thn\ interpretation is different. AEA is high in risk states due to chronic ARA release, which is the major problem in SZ. When ARA release becomes excessive, it overwhelms the capacity of AEA synthesis enzymes (FAAH reverse reaction, NAPE-PLD, others), and possibly of ARA release due to depletion, leading to a small decrease. At the same time, excessive ARA clogs the membrane and severely impairs its channels and receptors. It is this event that induces a conversion to full psychosis in the reported cases. 

Some support for this view is given by the decreased NAPE-PLD in FEP blood (see below). 

\smallest{FAAH}
FAAH activity is significantly increased in patient brains \cite{muguruza2019end} (without mRNA changes), with clinical remission associated with significantly decreased FAAH mRNA \cite{deMarchi2003end}.  
In addition, FEP patients show increased FAAH after controlling for cannabis use \cite{bioque2013per} (here there was a negative correlation between FAAH and symptoms before controlling for cannabis, which can be explained by the fact that exogenous cannabis reduces the activity of the ECB system \cite{morgan2013cer}.) 
These data accord with increased AEA, which is produced by FAAH via the reverse reaction when ARA levels are high. 

An association of SZ with NAPE-PLD and FAAH mutations has also been reported \cite{si2018ass}. 

\smallest{NAPE, NAPE-PLD}
Increased frontal cortex NAPE and decreased NAPE metabolites (N-acylethanolamines) were reported\cite{wood2019tar}. 
Accordingly, significantly decreased NAPE-PLD (the direct AEA synthesis enzyme) was reported in blood mononuclear cells in FEP \cite{bioque2013per}. 
%

\smallest{DAGL, MAGL}
One paper reported decreased DAGL (2AG synthesis) in FEP blood cells (before and after controlling for cannabis use), and increased MAGL (2AG degradation, only after controlling for cannabis) \cite{bioque2013per}. One paper reported no MAGL mRNA changes \cite{muguruza2019end}. 

\smallest{CB1 expression} 
Higher CB1 expression was reported in several papers, mainly using postmortem brain autoradiography but also PET and blood\cite{dean2001stu, zavitsanou2004sel, newell2006inc, wong2010qua, jenko2012bin, ceccarini2013inc, volk2014rec, chase2016cha, tao2020can, chou2023ter}. 
CB1 PET binding was higher in unmedicated than medicated patients (10\% vs 5\%) \cite{ceccarini2013inc}. 
One paper reported higher CB1 PET binding in paranoid SZ but not other SZ types \cite{dalton2011par}. 
An increase was reported only in patients who commited suicide, with a decrease in others \cite{tao2020can}. 

As predicted by \thn , downregulated CB1 (protein, mRNA, or ligand binding), which is a natural result of chronic signaling, is also frequently reported \cite{eggan2008red, uriguen2009imm, eggan2010can, ranganathan2016red, borgan2019viv, muguruza2019end, tao2020can, chou2023ter}. 
Decreased PFC CB1 was seen only in medicated patients \cite{uriguen2009imm}, supporting \desensit. 
CB1 was significantly higher in excitatory boutons, but lower in IINs \cite{chou2023ter}, again supporting \desensit, which is expected to be faster in IINs due to their higher activation level. 
A greater decrease was somewhat associated with higher symptom severity in males \cite{borgan2019viv}. 
Two papers reported no changes in superior temporal gyrus \cite{deng2007no} and anterior cingulate cortex \cite{koethe2007exp}. 

Recall that many of the negative effects of chronic ECB synthesis are \indep\ of the CB1 receptor, made via direct I-PUFA effects on ion channels, \icell\ \catwo, and membranes. Thus, chronic I-PUFAs can promote symptoms even when CB1 is downregulated. 

\smallest{Gi/o}
CB1 is coupled to Gi/o. Increased GPCR Go activation was reported in SZ \cite{jope1998sel, wallace1993tra}. 
A trend for higher basal CB1 coupling to Gi/o, indicating chronic activation, was also reported \cite{muguruza2019end}. 

\smallest{Genetic data} 
Patient blood shows decreased DNA methylation of the CB1 promoter, with a trend for increased expression \cite{dAddario2017pre}. 
As mentioned above, an association of SZ with NAPE-PLD and FAAH mutations was found \cite{si2018ass}. 
There is considerable data on CB1 mutations and SZ, but a recent thorough review found no consistent pattern \cite{navarro2022mol}. 

\smallest{Summary} 
ECB synthesis, signaling, and turnover are chronically increased in SZ. 

\subsection{Hyper\xtblty: neuromodulators, hormones, neurotransmitters}
\smallest{Neuromodulators: DA, NEP, SER, ACh}
The important neuromodulators of the brain, DA, NEP, SER, and ACh (which is also a neurotransmitter) are released by neurons located in specific nuclei. DA, NEP and SER neurons are regulated by ECBs \cite{covey2017end, carvalho2012can, haj2011mod}, which increases \spont\ firing and agent release. Thus, the effects of ECBs on neuromodulator release is similar to their effects in general. 
Indeed, there is direct evidence for slightly increased DA and NEP levels in SZ \cite{maia2017int, romer2023bio}. 
With respect to SER, there is only indirect evidence. 
The loudness dependence of auditory evoked potentials (LDAEP) is a measure of amplitude changes of auditory evoked electroencephalogram potentials in primary auditory cortex, and is known to be reduced by SER \cite{juckel2015ser}. LDAEP is reduced in SZ patients, which indicates higher cortical SER levels \cite{juckel2015ser}. 

The situation regarding ACh is more complex. On one hand, axonal inputs to basal forebrain ACh neurons express some CB1, and ACh neurons express FAAH \cite{harkany2003com}. 
On the other hand, CB1 agonists induce a prolonged inhibition of ACh release \cite{tzavara2003bip}, and ECBs clearly oppose ACh receptor signaling (see above). 
This makes sense, since there is substantial co-release of ACh and GABA \cite{saunders2015cor}, and ECBs inhibit GABAergic neurons to suppress coordinated network activity. 

Three ACh-related facts support \thn . 
First, there is widespread reduction in ACh receptor expression and signaling in SZ \cite{gibbons2016cho}.
Second, CB1 agonists induce cognitive deficits via decreased ACh \cite{robinson2010win, navakkode2014pha}. 
Third, SZ patients show increased prevalence of smoking \cite{tandon2008sch}, 
which is most likely a form of self-treatment. 


 
\smallest{Hormones: prolactin (PRL), cortisol}
PRL levels are increased in SZ, and this has been traditionally viewed as being due to the anti-DA effect of \antip\ medication. However, PRL levels are also greatly increased in first-episode, drug-naive patients \cite{gonzalez2016pro} and in high-risk people who converted to psychosis \cite{labad2015str}. 
This PRL increase supports \thn , because PLA2 increases PRL secretion \cite{grandison1984sti, ross1988dyn}. In fact, many known agents that stimulate PRL release, such as thyrotropin-releasing hormone, neurotensin, and angiotensin II, do so via PLA2 \cite{grandison1984sti, ross1988dyn}.  

Cortisol levels are also moderately increased in SZ, along with dysregulation of release rhythms \cite{pruessner2017neu, labad2015str}. This can be part of the epigenetic changes induced by stress, or more specifically be due to hyper\xtblty\ of neurons recruits GCs. 

\smallest{Neurotransmitters: GABA, glutamate}
The two main brain neurotransmitters, glutamate and GABA, are both affected in SZ, in the directions predicted by \thn . GABAergic IINs support coordinated and synchronous activity, and are normally suppressed by ECBs as part of their suppression of network activity. Chronic ECBs are predicted to induce a lasting decrease in GABA pathways. Indeed, CSF GABA is decreased in SZ \cite{romer2023bio}, 
and GABA synthesis enzymes, GABA transporters, and several GABAA subunits are decreased in patient postmortem dorsolateral PFC \cite{hashimoto2008alt}. These effects were not due to medication, since they did not occur in monkeys given chronic \antip s. 

Glutamate is the brain's main excitatory neurotransmitter, and predicted to be increased due to hyper\xtblty. Indeed, chronic spillover of glutamate from the synaptic cleft has been shown \cite{hammond2014evi}. This glutamate activates extrasynaptic receptors, producing a positive feedback loop that increases \xtblty. 
There is a well-known NMDAR hypofunction hypothesis of SZ \cite{friston2016dys}, but it is based on the psychosis-inducing effects of NMDARs \antag s. 

\smallest{Motor symptoms}
In addition to classical negative symptoms, which show decreased motor activity, SZ patients show motor symptoms that involve increased activity. In fact, catatonia in FEP patients mainly involves hyperkinesia, much more than hypokinesia \cite{peralta2010dsm}. 
Akathisia has been shown to be associated with worse prognosis \cite{cuesta2018mot}. 

\smallest{Neural \xtblty}
Patients with active paranoia show significantly increased resting cerebral blood flow activity in left \amg\ (and a trend in the right \amg) vs.\ other SZ patients or controls) \cite{pinkham2015amy}. 
SZ patients show higher tonic activity in right ventral striatum and bilateral \amg\ \cite{taylor2005neu}. 
A metareview found increased PET or fMRI activation in response to neutral stimuli in 20 out of 29 reports, including in the \amg\ (10), PFC (14), anterior cingulate cortex (6), and hippocampal formation (6) \cite{potvin2016emo}. 

Medicated SZ patients exhibit faster (and accurate) imagery, despite their working memory deficit \cite{matthews2014vis}. This supports priming (increased \xtblty) of high-level sensory experience paths. 

\subsection{Coordinated network activity}
There are several accepted measures of coordinated network activity, and all are reduced in SZ. 

\smallest{Event-related potentials (ERPs)}
ERPs measure the electroencephalography or magnetoencephalography responses to stimuli. ERPs include negativities and positivities at relatively fixed post-stimulus delays (e.g., the P300, N400), and some composite measures (e.g., the mismatch negativity). Shorter-term ERPs measure the brain's immediate responses to sensory inputs, while longer-term ERPs are a standard measure for the capacity of the brain to engage in coordinated top-down activity. 

ERPs, especially top-down ones, are reduced in SZ \cite{bodatsch2015for, berkovitch2018imp}. Anticipatory ERPs such as the contingent negative variation and the stimulus preceding negativity are reduced as well \cite{wynn2010imp}. 

\smallest{Oscillations}
Normal cortical execution involves electroencephalography oscillations at various frequency bands. Oscillations and synchrony are disrupted in SZ \cite{skosnik2016its, uhlhaas2015osc}. 

\smallest{Cortical activity}
In addition to these measures, a well-known result in SZ, sometimes termed hypofrontality, is that of decreased PFC activity in resting and cognitive challenge conditions \cite{tandon2008sch}. 
In addition, there is hypoperfusion in wide brain areas \cite{lopes2015ang}. 

\smallest{Synapses}
Synapses are essential for brain execution. There is clear evidence for reduced synaptic density and for synaptic dysfunction in SZ, including in ion channels, exocytosis, glutamate receptors, postsynaptic density, and adhesion molecules \cite{misir2023syn}. 

\smallest{Brain volume}
Another well-supported result in SZ is reduced total brain volume, with larger third and lateral ventricular spaces \cite{tandon2008sch}. This is not a direct index of coordinated activity, but it indicates impaired \plast\ and development. 

\subsection{Summary}
\thn\ enjoys overwhelming support from virtually all bodies of evidence gathered about SZ in the last 50 years. On the agent side, there is increased I-PUFA and decreased E-PUFA synthesis and signaling, increased ECB (especially AEA) synthesis, and increased CB1 activation and de\sensit. On the effect side, there is clear neural hyper\xtblty\ coupled with decreased coordinated cortical activity. These effects induce the symptoms of SZ as explained above. 

\refsection{Treatment}{treat}


Here I discuss novel and existing treatment directions, including iPLA2 suppression, dietary om3, CB1 antagonism, NAC, and \antip s. I explain the mechanisms of \antip s and discuss their upsides and downsides. 

\smallest{Reducing iPLA2}
According to \thn , SZ involves excessive PLA2 activity, mainly iPLA2. Hence, the most promising drug-based direction is to reduce iPLA2 activity. 
Several families of potent and highly selective small molecule iPLA2 inhibitors have been designed \cite{mouchlis2016dev, dedaki2019bet, smyrniotou2017oxo}. None of them has been tested on humans. The published papers provide extensive explanations of design considerations and techniques. 

Alternatively, there are non-selective PLA2 inhibitors that are readily available for treating humans. Anti-malaria drugs such as chloroquine, quinacrine, hydroxychloroquine, and quinine have been used by many millions of people for decades, and are PLA2 inhibitors, with efficacy ranking at this order \cite{lu2001dif}. 
These drugs cross the blood-brain barrier \cite{dubin1982pha}, and are widely available and inexpensive.  
SZ prevalence is indeed lowest in areas where anti-malaria drugs are used \cite{charlson2018glo} (although this datum is not strong). 
Thus, anti-malaria drugs are candidates for treating SZ. 

This proposal might seem paradoxical, because quinacrine and related drugs can induce psychosis \cite{nevin2016psy}. 
This is not surprising, since, in addition to its effect on PLA2, quinacrine inhibits muscarinic \cite{oDonnell1991mus} and nicotinic \cite{tamamizu1995mut} ACh receptors. In addition, when given to people with normal PLA2 activity (i.e., people without SZ), quinacrine can induce excessive I-PUFA suppression, which can cause excessive activation of the ACh system, followed by its \desensit. This can disconnect the neural network and increase \spont\ spiking to induce psychosis in the same manner as ECBs and muscarinic \antag s do. 

In spite of this, there are arguments to support the usage of anti-malaria drugs for treating SZ. First, psychosis is a rare adverse event. Among 30,000 soldiers treated for malaria, psychosis occurred in only 0.1-0.4\% \cite{lidz1946tox}. 
Second, psychosis overwhelmingly occurred when using high drug doses, higher than those used for normal treatment \cite{newell1946tox}. 
Doses resulting in plasma levels of less than 18 microgram per 100 cubic cm did not yield any in mental or intellectual impairment \cite{lidz1946tox}. 
Since the PLA2 upregulation in SZ is a small-scale phenomenon, relatively small inhibitor doses would be needed in SZ, which almost completely avoids risk. 
Third, the reports of anti-malaria drugs causing psychosis are with people without SZ. As explained above, \thn\ predicts that PLA2 inhibition could yield psychosis in healthy people, but when used with people with SZ, their effect would be to normalize an impaired state, not impair a healthy state. 
Finally, quinacrine is a structural antecedent of the phenothiazines, and its structure is similar to that of chlorpromazine \cite{kitagawa2021ant}, the first antipsychotic drug\footnote{In light of \thn , it is hypothesized here that the benefits of chlorpromazine are due to a PLA2 effect more than to other discussed mechanisms. Indeed, it is a highly potent inhibitor of cPLA2 \cite{vadas1986pot}.}. 
Chlorpromazine is effective against positive symptoms, but has undesired side effects. The fact that a highly similar molecule has been shown to be effective in SZ provides some validation with respect to PLA2 inhibitors and quinacrine. 

Even if not used directly, non-selective iPLA2 inhibitors constitute a promising direction to pursue at present, since it is possible that relatively small modifications could oppose their ACh suppression and make them more selective towards iPLA2 than to cPLA2. 

\smallest{Dietary PUFA}
According to \thn , ARA's role in SZ is most likely much larger than that of DHA. As a result, reducing the om6/om3 ratio via diet is probably a good idea, certainly for patients who show om3 deficiency. Indeed, the evidence section above summarized evidence that om3 may have a protective effect, and that dietary om3 supplementation might be beneficial at the initial stages of the disease. The lack of benefit at later stages may be related to the supplementation being adjunctive to \antip s. 

\smallest{CB1 \antag s}
Rimonabant and other related CB1 \antag s were tested in long trials for reducing obesity. They turned out to be moderately effective for weight loss, but to increase the risk of depression and anxiety \cite{christensen2007eff}. For this reason, rimonabant was withdrawn from the market. 
Since ECBs are recruited to facilitate the early stages of stress termination, mood symptoms can be expected when ECB signaling is reduced in healthy people. In addition, CB1 antagonism may yield accumulation of ARA and increased \spont\ \xtblty. 

Nonetheless, in SZ, the ECB system is hyperactive, and CB1 antagonism may normalize it to some extent, especially in patients with increased CB1 expression. Indeed, in a small clinical trial using rimonabant for the treatment of obesity in SZ, patients showed a significantly decreased psychiatric score and anxiety. There were no adverse effects \cite{kelly2011eff}, but the trial was terminated early due to the drug's withdrawal. 
In another trial, rimonabant had no effect on SZ symptoms and no adverse effects \cite{meltzer2004pla}. 
Similarly, a trial using co-treatment with the CB1 \antag\ AVE1625 and atypical \antip s was terminated early due to insufficient benefit \cite{sanofi2009SZ}. 
CB1 antagonism does not address the root problem of excessive ARA in SZ, since it can promote ARA accumulation. This might explain the lack of efficacy in these trials. 

Cannabidiol is a cannabis component that counters the psychotic effects of cannabis, possibly by opposing CB1 signaling \cite{iseger2015sys}. It can be expected to have effects in a similar direction to CB1 \antag s, but weaker. 

\smallest{N-acetylcysteine (NAC)}
Based on the assumption that NAC might be beneficial in SZ by reducing \oxis, several clinical trials have been conducted, with positive results. In a systematic review and network meta-analysis, adjunctive NAC, but not om3 PUFAs, folic acid, vitamin B12, or vitamin D, was found to significantly reduce symptoms vs.\ placebo \cite{xu2022eff}. 
In my view, the main mechanism here is not \oxis\ but PGs. Paracetamol reduces pain by suppressing COX, and NAC is routinely used against paracetamol overdose, with common skin rash, reddening, and itch side effects. This implies that NAC probably enhances PG synthesis, which should be beneficial in SZ due to channeling of ARA utilization away from I-PUFAs. 

One paper reported that ultraviolet radiation stimulated cPLA2 synthesis, which was correlated with increased ARA and PGE2, and that this stimulation was prevented by NAC \cite{chen1996oxi}. 
This result contradicts the PG assumption above, but suppression of cPLA2 synthesis could be \indep ly beneficial in SZ. 

\smallest{\Antip s}
Today, SZ is treated with so-called \antip s, which are all D2 \antag s. Atypical \antip s are also SER2a \antag s. Like CB1, D2 is a GPCR coupled to Gi/o, and their brain distributions have a moderate overlap, especially in the basal ganglia striatum, PFC \cite{quintana2019D2} and \amg\ \cite{deLaMora2012dis}. 
D2 agonists greatly increase AEA release and suppress its degradation \cite{giuffrida1999dop, patel2003dif, centonze2004cri}. 
Thus, antagonizing D2 reduces the excessive AEA present in SZ. 

SER2a is excitatory and has a wide brain distribution. In particular, it is presynaptically expressed in thelamocortical neurons, and postsynaptically expressed in layer 5 PFC pyramidal neurons \cite{barre2016pre}, so antagonizing it reduces the \xtblty\ of many neurons participating in sensory perception. 
These data explain the benefit of these drugs on positive symptoms. 

There is evidence that these drugs improve patient ECB and PUFA state.
Olanzapine-induced remission was associated with decrease in blood AEA and FAAH mRNA \cite{deMarchi2003end}. 
Patients treated with typical (but not atypical) \antip s did not show the excessive CSF ARA shown in FEP and atypical medicated patients \cite{giuffrida2004cer}. 
Eight week treatment reduced higher serum AEA and increased the lower serum COX products (PGE2 and others) \cite{wang2018alt}. 
Treatment partially normalized decreased membrane ARA and DHA \cite{mcnamara2007abn, li2022red}, 
and decreased PLA2 activity \cite{gattaz1987inc, tavares2003inc}. 
The odds that treatment improves negative symptoms was higher with high basal iPLA2 activity \cite{smesny2011pho}. 
\Antip s decrease iPLA2 gene expression in patients \cite{kerr2013ant}. 
Clozapine upregulated COX2 mRNA in human \astc s \cite{yuhas2022clo}, which might be beneficial by channeling ARA to E-PUFAs rather than I-PUFAs. 
In addition, one preclinical study found that chronic clozapine decreased rat plasma ARA, brain COX and PGE2, and the incorporation of ARA into the brain \cite{modi2013chr}. 
Other studies found that D2 \cite{felder1991tra, piomelli1991dop} and SER2 \cite{felder1990ser} potentiate PLA2, 
and that olanzapine decreases LOX5 activating protein \cite{dzitoyeva2013lox}. 

However, there is also data showing that \antip s may exacerbate patient I-PUFA state. \Antip s activate SREBP genes \cite{raeder2006sre}, which stimulate iPLA2 mRNA and protein \cite{ramanadham2015cal}. 
They also upregulate the desaturase enzymes that participate in the production of ARA and DHA from their precursors \cite{polymeropoulos2009com}. 
Chronic risperidone, olanzapine, and haloperidol increased ARA and DHA in red blood cells (quetiapine increased only DHA) \cite{mcnamara2011dif}. 
30 days of clozapine increased iPLA2 (but not other PLA2) mRNA, protein, and activity, and decreased COX activity and PGE2 in rats \cite{kim2012eff}. 
30 days of olanzapine increased frontal cortex iPLA2 and COX1 mRNA, and decreased brain PGE2 and ARA brain incorporation \cite{cheon2011chr}. 
Seizures are a frequent adverse event of \antip s \cite{wu2016com}, which can be explained via increased \xtblty\ or upregulated PG (recall that PGs are highly excitatory). 

CB1-D2 heteromers are common, and co-stimulation of the receptors promote heteromer formation and reverses CB1 signaling to induce cAMP \cite{kearn2005con}. Since CB1 is chronically stimulated in SZ, stimulation of D2 should be beneficial in SZ by reversing CB1 signaling. In this view, \antip s may enhance chronic CB1 signaling. 
However, after treatment of cells expressing CB1 and D2 with a CB1 agonist, haloperidol switched CB1 to inducing cAMP \cite{bagher2016ant}. These conflicting data do not allow firm conclusions. 

SZ patients show increased CB1-SER2a heteromers and inhibition of cAMP (i.e., classic CB1 signaling) in the olfactory neuroepithelium, due to a switch of SER2a signaling from Gq to Gi \cite{guinart2020alt}. Co-stimulation additively decreased cAMP in non-cannabis users treated with clozapine. Here, \antip s enhance CB1-like signaling, which would worsen patient state. 

At the functional level, SER2a promotes active coping in stress and enhances learning and memory \cite{carhart2017ser}, so antagonizing it could be harmful if it does not directly target the disorder causes. 

Importantly, atypical \antip s have strong anticholinergic properties \cite{chew2008ant}, and many of the detrimental effects of ECBs in SZ are due to opposition to ACh. 
A very common adverse effect after months or years of treatment is a movement disorder called tardive dyskinesia (TD), which has many similarities with Parkinsonism \cite{bordia2016str}. 
TD is usually treated using muscarinic \antag s (although improvement after their discontinuation has also been reported) \cite{bordia2016str}. 
It has been shown in mice that catalepsy (the mouse parallel of TD) depends on D2 expression in cholinergic neurons and a subsequent excessive ACh in the striatum \cite{kharkwal2016PD}, probably leading to \desensit\ of ACh receptors. This may imply that ECB- and \antip-induced ACh damage combine in SZ to exacerbate long-term patient state. 

There is \indep\ evidence that long-term usage of \antip s may be detrimental. 
A 20-year followup study found that patients not on drugs after the first two years have a higher chance of recovery \cite{harrow2022twe}. 
However, a recent clinical trial reported that gradual \antip s dose reduction over two years increased the occurrence of adverse psychotic events without having any effect on other symptoms or general functioning \cite{moncrieff2023ant}. 

Drugs probably get desensitized. 
The response to drugs, measured via phospholipids (PE, lysoPC) and neurosteroids, diminished after 6 months \cite{schmitt2001eff} or 5 years \cite{cai2022dim}. 
In a clinical trial, FEP patients improved similarly with drugs or placebo after the first 6 months \cite{francey2020psy}. 

In summary, \thn\ can explain the mechanisms through which \antip s yield beneficial results on positive symptoms, and why they have only a weak effect on other symtpoms. These drugs affect PLA2 and its products, with most of the short-term effects reported in SZ being positive. However, these reports are inconsistent, while negative long-term effects are also reported, implying that chronic \antip s might be detrimental via PLA2 for some patients. 

\smallest{AChR drugs} 
There are ongoing efforts to develop muscarinic agonists (positive allosteric modulators) for SZ (e.g., by Karuna, Cerevel, MapLight, and Neumora). As explained above, this direction is biologically valid, since ECBs suppress these receptors. However, such drugs address only part of the problem (and weakly so), so I do not expect them to provide substantial benefit to patients across the symptom spectrum when used alone. 

\smallest{Biomarkers} 
According to \thn , increased blood or brain iPLA2 (or any PLA2) activity, increased CSF ECB (mainly AEA) levels, and decreased membrane ARA and DHA in blood or brain cells, all indicate a high risk of SZ, especially when combined with reduced PGs (mainly PGE2). Indeed, AEA, ARA and oleoylethanolamine \cite{wang2018alt}, and PE, PC, LA and some other lipids \cite{song2023pot} allow very good  classification accuracy. 
Since blood PLA2, ARA and PGs can easily be measured, this is the most promising biomarker for identifying high risk and for monitoring treatment effects. 

\refsection{Discussion}{disc}
SZ is a major brain disorder and has been a mystery to humankind for a long time. 
This paper presented the PUFA theory of SZ (\thn ), in which chronic synthesis of I-PUFAs (ARA, DHA, ECBs (mainly AEA), Lkts), usually stemming from stress occurring in sensitive periods, causes the symptoms of SZ. \thn\ is supported by a large body of evidence showing excessive I-PUFA synthesis, elevated ECB system activity, hyper\xtblty\ of neuromodulators and neurotransmitters, and the effects of \antip s. The theory explains the benefits and limitations of current treatment and implies novel treatment directions. 

\smallest{Related theories}
\thn\ is a complete theory, covering SZ etiology, symptoms, pathology, risk factors, and treatment. It seems to be the first theory of SZ that provides a mechanistic account of SZ symptoms\footnote{Note: I used the word `disconnect' when describing the effects of ECBs on the synthesizing cell. To prevent any confusion, it should be noted that \thn\ has absolutely no relationship with the `dysconnection hypothesis of SZ', which discusses the effect of impaired DA on synapses \cite{friston2016dys}.}. 
It is also the first to recognize the connection between ECBs and the strong evidence involving PUFA in SZ.

The links between SZ and ECBs are well-known and have been extensively reviewed \cite{desfosses2010end, leweke2017put}. 
Such work highlights the effect of cannabis on SZ risk, the fact that cannabis can induce psychotic symptoms, and changes to the ECB system (mainly to CB1 density) in SZ.
%
I found only one paper that goes further to propose that hyperactivity of the ECB system is involved in the pathogenesis of SZ \cite{muller2008can}. 
However, even this paper does not propose any mechanistic account, and its authors justly present it as moving towards an ECB hypothesis of SZ rather than a theory.

Similarly, the link between SZ and PUFAs (including ARA and PLA2) has already been recognized, as attested by the clinical trials using om3 supplementation. However, I am not aware of any previous proposal for how PUFA alterations might induce the symptoms of SZ, nor of any recognition of the relationship between the PUFA and the ECB evidence in SZ. 

One of the first hypotheses for the cause of SZ was that it is a PG deficiency disorder \cite{horrobin1977sch}. However, this hypothesis was based on a small set of observations, not on a mechanistic account. In sharp contrast to \thn , it concluded with the proposal that ARA should help patients. 
\smallest{Dopamine}
The DA hypothesis figures prominently in all current reviews and textbooks of SZ. It is based on the perceived efficacy of D2 \antag s, the psychotic effects of amphetamine, and the moderately increased DA levels in SZ. However, no detailed proposal has been made as to how DA induces SZ symptoms. In \thn , increased DA levels are a consequence of the uncontrolled \xtblty\ induced by ECBs when acting on DA neurons. However, ECBs induce a similar effect on almost all other cell types in the brain as well. The effect of D2 \antag s can be explained via the partial overlap in distribution and signaling between D2 and CB1, while the effect of amphetamine can be explained via its action on D2. I conclude that dopamine does not have any special role in SZ. 

\smallest{The role of iPLA2}
The ARA used for I-PUFA synthesis is mainly released by iPLA2. In this sense, this paper could have been titled ``the iPLA2 theory of SZ''. However, \thn\ does not claim that iPLA2 is the sole cause of SZ. At its root, SZ is caused by persistent expression changes following prolonged stress. These most likely lead to upregulated iPLA2, but can also involve a wide range of additional effects (e.g., suppression of PG pathway genes, enhancement of AEA and Lkt genes, increased constitutive CB1 activity, etc). iPLA2 upregulation itself can occur due to \epig\ alterations or mutations to its gene, but can also involve upstream genes (e.g., genes regulating store-operated calcium channels, calcium pumps, \er, \mitoch, etc). These alterations can be relatively small and difficult to identify, at least using current technology. 

In the acronym `iPLA2', the letter `i' can conveniently denote `\icell' rather than `\indep'. The description `intracellular calcium store-induced PLA2' is a more appropriate name for this enzyme than `calcium-\indep\ PLA2'. 

\smallest{Cannabis}
\thn\ does not claim that SZ is directly caused by cannabis use. However, exogenous cannabis activates ECB signaling and hence greatly increases risk in susceptible individuals. Moreover, prolonged cannabis use mimics some aspects of chronic stress, so can induce prolonged symptoms, especially if the exposure occurs in sensitive developmental periods. 


\smallest{Symptoms}
In \thn , positive symptoms are mainly due to excessive \xtblty, while non-positive symptoms are mainly due to network disconnect. This is one reason why drugs that counter the former are not beneficial for the latter. 

It is not the case that `lack of affect' negative symptoms in SZ occur because SZ patients do not experience emotions. On the contrary, their capacity for experiencing hedonic emotions is intact \cite{cohen2010emo}. These symptoms occur because controlled access to emotion representations and responses is impaired \cite{kring2010emo}. 
The problem mainly manifests in emotion anticipation, and is not limited to hedonic emotions, occurring also with respect to losses \cite{yan2019ant}. 
Thus, affect symptoms can be viewed as a subtype of cognitive symptoms, both being due to network disconnect. 


\smallest{Age of onset}
SZ is usually diagnosed during adolescence or some time later, with the eruption of psychosis. Juvenile psychosis exists, but it is rare (incidence less than 0.04\% \cite{driver2013chi}), 
while the other types of symptoms can be present throughout the person's life and are commonly present prodromally. 

The relatively late appearance of psychosis compared to other symptoms may be explained by the fact that \adol\ is a period with massive cortical \plast. In particular, GC (cortisol) is one of the major recruiters of ECBs, and \adol\ is when the expression of the GC receptor is at its peak \cite{sinclair2011dyn}.
Thus, \adol\ involves a strong enhancement of the ECB system, which drives it above the psychosis threshold in people who already exhibit the capacity for chronic ECB activity. 


\smallest{Sex differences}
SZ is diagnosed earlier in males, and women show increased incidence following menopause \cite{markham2012sex}. 
This probably happens because estrogen, although it recruits ECBs \cite{gorzalka2012min}, is neuroprotective via diverse mechanisms including \mitoch, anti-\infl, anti-\oxis, synaptic \plast, ACh, growth factors, the microvasculature, and the blood-brain barrier \cite{engler2017est}. 
Estrogen is elevated earlier in females than \testos\ in males, delaying disease onset. 
 
Testosterone is neuroprotective as well, and indeed, the sex difference in SZ diminishes with age. Testosterone is reported decreased or unchanged in SZ \cite{markham2012sex}, with the decrease most likely being due to increased PRL and to GC opposition of gonadal steroids. 


\smallest{Future work}
Large parts of the discussion here apply to bipolar disorder and depression as well. 
Some of the analysis here can be used to improve our understanding of immunity-related diseases. 
The animal models used today for SZ are not valid, but it is possible to develop valid ones. 
These topics and others will be discussed elsewhere. 


\smallest{Theory predictions}
The synthesis and signaling mechanisms of ARA and its products are well-known. This paper has amassed strong evidence for increased I-PUFAs, disconnected network, and hyper\xtblty\ in SZ, and for their capacity of inducing SZ symptoms. Thus, most of the major aspects of \thn\ have already been validated. The areas that most lack evidence are the \epig\ or genetic mechanisms that upregulate the PLA2/ARA paths. A theory prediction is that such mechanisms exist. It might be difficult to find them, as noted above. 

The ultimate theory prediction is that normalizing the I-PUFA synthesis paths would improve patient symptoms. Normalization might not be able to undo all of the brain damage done until treatment has started, but it might alleviate it, hopefully to practically unnoticeable levels.



\bibliographystyle{vancouver} 
\bibliography{SZ,RRR-ALS,RRR-AZ} 

\begin{thebibliography}{100}

\bibitem{tandon2008sch}
Tandon R, Keshavan MS, Nasrallah HA.
\newblock Schizophrenia, ``just the facts'': what we know in 2008: part 1:
  overview.
\newblock Schizophrenia research. 2008;100(1-3):4--19.

\bibitem{fivsar2022bio}
Fi{\v{s}}ar Z.
\newblock Biological hypotheses, risk factors, and biomarkers of schizophrenia.
\newblock Progress in Neuro-Psychopharmacology and Biological Psychiatry.
  2022;p. 110626.

\bibitem{owen2016sz}
Owen MJ, Sawa A, Mortensen PB.
\newblock Schizophrenia.
\newblock Lancet. 2016;388:86--97.

\bibitem{leweke2017put}
Leweke FM, Rohleder C.
\newblock Putative role of endocannabinoids in schizophrenia.
\newblock In: The Endocannabinoid System. Elsevier; 2017. p. 83--113.

\bibitem{hoen2013red}
Hoen WP, Lijmer JG, Duran M, Wanders RJ, van Beveren NJ, de~Haan L.
\newblock Red blood cell polyunsaturated fatty acids measured in red blood
  cells and schizophrenia: a meta-analysis.
\newblock Psychiatry research. 2013;207(1-2):1--12.

\bibitem{muller2008can}
M{\"u}ller-Vahl KR, Emrich HM.
\newblock Cannabis and schizophrenia: towards a cannabinoid hypothesis of
  schizophrenia.
\newblock Expert Review of Neurotherapeutics. 2008;8(7):1037--1048.

\bibitem{cortes2015psy}
Cortes-Briones JA, Cahill JD, Skosnik PD, Mathalon DH, Williams A, Sewell RA,
  et~al.
\newblock The psychosis-like effects of $\Delta$9-tetrahydrocannabinol are
  associated with increased cortical noise in healthy humans.
\newblock Biological psychiatry. 2015;78(11):805--813.

\bibitem{marconi2016met}
Marconi A, Di~Forti M, Lewis CM, Murray RM, Vassos E.
\newblock Meta-analysis of the association between the level of cannabis use
  and risk of psychosis.
\newblock Schizophrenia bulletin. 2016;42(5):1262--1269.

\bibitem{crosby2012int}
Crosby KM, Bains JS.
\newblock The intricate link between glucocorticoids and endocannabinoids at
  stress-relevant synapses in the hypothalamus.
\newblock Neuroscience. 2012;204:31--37.

\bibitem{macdonald2009we}
MacDonald AW, Schulz SC.
\newblock What we know: findings that every theory of schizophrenia should
  explain.
\newblock Schizophrenia bulletin. 2009;35(3):493--508.

\bibitem{varese2012chi}
Varese F, Smeets F, Drukker M, Lieverse R, Lataster T, Viechtbauer W, et~al.
\newblock Childhood adversities increase the risk of psychosis: a meta-analysis
  of patient-control, prospective-and cross-sectional cohort studies.
\newblock Schizophrenia bulletin. 2012;38(4):661--671.

\bibitem{dAddario2013epi}
D'Addario C, Di~Francesco A, Pucci M, Finazzi~Agr{\`o} A, Maccarrone M.
\newblock Epigenetic mechanisms and endocannabinoid signalling.
\newblock The FEBS journal. 2013;280(9):1905--1917.

\bibitem{meccariello2020epi}
Meccariello R, Santoro A, D'Angelo S, Morrone R, Fasano S, Viggiano A, et~al.
\newblock The epigenetics of the endocannabinoid system.
\newblock International journal of molecular sciences. 2020;21(3):1113.

\bibitem{scott2020int}
Scott MR, Meador-Woodruff JH.
\newblock Intracellular compartment-specific proteasome dysfunction in
  postmortem cortex in schizophrenia subjects.
\newblock Molecular psychiatry. 2020;25(4):776--790.

\bibitem{serhan2014pro}
Serhan CN.
\newblock Pro-resolving lipid mediators are leads for resolution physiology.
\newblock Nature. 2014;510(7503):92--101.

\bibitem{maccarrone2008ana}
Maccarrone M, Rossi S, Bari M, De~Chiara V, Fezza F, Musella A, et~al.
\newblock Anandamide inhibits metabolism and physiological actions of
  2-arachidonoylglycerol in the striatum.
\newblock Nature neuroscience. 2008;11(2):152--159.

\bibitem{glasner2014meth}
Glasner-Edwards S, Mooney LJ.
\newblock Methamphetamine psychosis: epidemiology and management.
\newblock CNS drugs. 2014;28:1115--1126.

\bibitem{walentiny2010kap}
Walentiny DM, Vann RE, Warner JA, King LS, Seltzman HH, Navarro HA, et~al.
\newblock Kappa opioid mediation of cannabinoid effects of the potent
  hallucinogen, salvinorin {A}, in rodents.
\newblock Psychopharmacology. 2010;210:275--284.

\bibitem{hu2015glut}
Hu W, MacDonald ML, Elswick DE, Sweet RA.
\newblock The glutamate hypothesis of schizophrenia: evidence from human brain
  tissue studies.
\newblock Annals of the New York Academy of Sciences. 2015;1338(1):38--57.

\bibitem{ross2012ste}
Ross DA, Cetas JS.
\newblock Steroid psychosis: a review for neurosurgeons.
\newblock Journal of Neuro-oncology. 2012;109:439--447.

\bibitem{barre2016pre}
Barre A, Berthoux C, De~Bundel D, Valjent E, Bockaert J, Marin P, et~al.
\newblock Presynaptic serotonin {2A} receptors modulate thalamocortical
  plasticity and associative learning.
\newblock Proceedings of the National Academy of Sciences.
  2016;113(10):E1382--E1391.

\bibitem{wu2010req}
Wu CS, Zhu J, Wager-Miller J, Wang S, O'Leary D, Monory K, et~al.
\newblock Requirement of cannabinoid {CB1} receptors in cortical pyramidal
  neurons for appropriate development of corticothalamic and thalamocortical
  projections.
\newblock European Journal of Neuroscience. 2010;32(5):693--706.

\bibitem{jakab2000SER}
Jakab RL, Goldman-Rakic PS.
\newblock Segregation of serotonin {5-HT2A} and {5-HT3} receptors in inhibitory
  circuits of the primate cerebral cortex.
\newblock Journal of Comparative Neurology. 2000;417(3):337--348.

\bibitem{bazinet2014PUFA}
Bazinet RP, Lay{\'e} S.
\newblock Polyunsaturated fatty acids and their metabolites in brain function
  and disease.
\newblock Nat Rev Neurosci. 2014;15(12):771--785.

\bibitem{wilkins2008gro}
Wilkins~Iii W, Barbour S, et~al.
\newblock Group {VI} phospholipases {A2}: homeostatic phospholipases with
  significant potential as targets for novel therapeutics.
\newblock Current Drug Targets. 2008;9(8):683--697.

\bibitem{barbour1999reg}
Barbour SE, Kapur A, Deal CL.
\newblock Regulation of phosphatidylcholine homeostasis by calcium-independent
  phospholipase {A}2.
\newblock Biochimica et Biophysica Acta (BBA)-Molecular and Cell Biology of
  Lipids. 1999;1439(1):77--88.

\bibitem{negre1996cha}
N{\`e}gre-Aminou P, Nemenoff RA, Wood MR, de~La~Houssaye B, Pfenninger K.
\newblock Characterization of phospholipase {A}2 activity enriched in the nerve
  growth cone.
\newblock Journal of neurochemistry. 1996;67(6):2599--2608.

\bibitem{rosa2009int}
Rosa AO, Rapoport SI.
\newblock Intracellular-and extracellular-derived {C}a2+ influence
  phospholipase {A}2-mediated fatty acid release from brain phospholipids.
\newblock Biochimica et Biophysica Acta (BBA)-Molecular and Cell Biology of
  Lipids. 2009;1791(8):697--705.

\bibitem{guo2010ind}
Guo C, Li J, Myatt L, Zhu X, Sun K.
\newblock Induction of {G}$\alpha$s contributes to the paradoxical stimulation
  of cytosolic phospholipase {A}2$\alpha$ expression by cortisol in human
  amnion fibroblasts.
\newblock Molecular Endocrinology. 2010;24(5):1052--1061.

\bibitem{ramanadham2015cal}
Ramanadham S, Ali T, Ashley JW, Bone RN, Hancock WD, Lei X.
\newblock Calcium-independent phospholipases {A2} and their roles in biological
  processes and diseases.
\newblock Journal of lipid research. 2015;56(9):1643--1668.

\bibitem{katona2012mul}
Katona I, Freund TF.
\newblock Multiple functions of endocannabinoid signaling in the brain.
\newblock Annual review of neuroscience. 2012;35:529--558.

\bibitem{sun2004bio}
Sun YX, Tsuboi K, Okamoto Y, Tonai T, Murakami M, Kudo I, et~al.
\newblock Biosynthesis of anandamide and {N}-palmitoylethanolamine by
  sequential actions of phospholipase {A2} and lysophospholipase {D}.
\newblock Biochemical Journal. 2004;380(3):749--756.

\bibitem{liu2006bio}
Liu J, Wang L, Harvey-White J, Osei-Hyiaman D, Razdan R, Gong Q, et~al.
\newblock A biosynthetic pathway for anandamide.
\newblock Proceedings of the National Academy of Sciences.
  2006;103(36):13345--13350.

\bibitem{tsuboi2018end}
Tsuboi K, Uyama T, Okamoto Y, Ueda N.
\newblock Endocannabinoids and related {N}-acylethanolamines: biological
  activities and metabolism.
\newblock Inflammation and Regeneration. 2018;38:1--10.

\bibitem{busquets2018cb1}
Busquets-Garcia A, Bains J, Marsicano G.
\newblock {CB1} receptor signaling in the brain: extracting specificity from
  ubiquity.
\newblock Neuropsychopharmacology. 2018;43(1):4--20.

\bibitem{maroso2016can}
Maroso M, Szabo GG, Kim HK, Alexander A, Bui AD, Lee SH, et~al.
\newblock Cannabinoid control of learning and memory through {HCN} channels.
\newblock Neuron. 2016;89(5):1059--1073.

\bibitem{goppelt1989glu}
Goppelt-Struebe M, Wolter D, Resch K.
\newblock Glucocorticoids inhibit prostaglandin synthesis not only at the level
  of phospholipase {A2} but also at the level of cyclo-oxygenase/{PGE}
  isomerase.
\newblock British journal of pharmacology. 1989;98(4):1287--1295.

\bibitem{ristimaki1996dow}
Ristim{\"a}ki A, Narko K, Hla T.
\newblock Down-regulation of cytokine-induced cyclo-oxygenase-2 transcript
  isoforms by dexamethasone: evidence for post-transcriptional regulation.
\newblock Biochemical Journal. 1996;318(1):325--331.

\bibitem{xia2012pro}
Xia D, Wang D, Kim SH, Katoh H, DuBois RN.
\newblock Prostaglandin {E2} promotes intestinal tumor growth via {DNA}
  methylation.
\newblock Nature medicine. 2012;18(2):224--226.

\bibitem{zhang2008end}
Zhang J, Chen C.
\newblock Endocannabinoid 2-arachidonoylglycerol protects neurons by limiting
  {COX}-2 elevation.
\newblock Journal of Biological Chemistry. 2008;283(33):22601--22611.

\bibitem{chang2004clo}
Chang WC, Parekh AB.
\newblock Close functional coupling between {C}a2+ release-activated {C}a2+
  channels, arachidonic acid release, and leukotriene {C}4 secretion.
\newblock Journal of Biological Chemistry. 2004;279(29):29994--29999.

\bibitem{finney2009leu}
Finney-Hayward TK, Bahra P, Li S, Poll CT, Nicholson AG, Russell RE, et~al.
\newblock Leukotriene {B4} release by human lung macrophages via receptor-not
  voltage-operated {C}a2+ channels.
\newblock European Respiratory Journal. 2009;33(5):1105--1112.

\bibitem{smani2003ca2}
Smani T, Zakharov SI, Leno E, Csutora P, Trepakova ES, Bolotina VM.
\newblock Ca2+-independent phospholipase {A}2 is a novel determinant of
  store-operated {C}a2+ entry.
\newblock Journal of Biological Chemistry. 2003;278(14):11909--11915.

\bibitem{brash2001ara}
Brash AR, et~al.
\newblock Arachidonic acid as a bioactive molecule.
\newblock The Journal of clinical investigation. 2001;107(11):1339--1345.

\bibitem{oz2006rec}
Oz M.
\newblock Receptor-independent actions of cannabinoids on cell membranes: focus
  on endocannabinoids.
\newblock Pharmacology \& therapeutics. 2006;111(1):114--144.

\bibitem{lagalwar1999ana}
Lagalwar S, Bordayo EZ, Hoffmann KL, Fawcett JR, Frey WH.
\newblock Anandamides inhibit binding to the muscarinic acetylcholine receptor.
\newblock Journal of Molecular Neuroscience. 1999;13:55--61.

\bibitem{varga2014ana}
Varga A, Jenes A, Marczylo TH, Sousa-Valente J, Chen J, Austin J, et~al.
\newblock Anandamide produced by {C}a2+-insensitive enzymes induces excitation
  in primary sensory neurons.
\newblock Pfl{\"u}gers Archiv-European Journal of Physiology.
  2014;466:1421--1435.

\bibitem{sharir2010pha}
Sharir H, Abood ME.
\newblock Pharmacological characterization of {GPR55}, a putative cannabinoid
  receptor.
\newblock Pharmacology \& therapeutics. 2010;126(3):301--313.

\bibitem{boland2008pol}
Boland LM, Drzewiecki MM.
\newblock Polyunsaturated fatty acid modulation of voltage-gated ion channels.
\newblock Cell biochemistry and biophysics. 2008;52:59--84.

\bibitem{vreugdenhil1996pol}
Vreugdenhil M, Bruehl C, Voskuyl R, Kang J, Leaf A, Wadman W.
\newblock Polyunsaturated fatty acids modulate sodium and calcium currents in
  {CA1} neurons.
\newblock Proceedings of the National Academy of Sciences.
  1996;93(22):12559--12563.

\bibitem{bouzat1993eff}
Bouzat C, Barrantes F.
\newblock Effects of long-chain fatty acids on the channel activity of the
  nicotinic acetylcholine receptor.
\newblock Receptors \& channels. 1993;1(3):251--258.

\bibitem{nabekura1998fun}
Nabekura J, Noguchi K, Witt MR, Nielsen M, Akaike N.
\newblock Functional modulation of human recombinant $\gamma$-aminobutyric acid
  type {A} receptor by docosahexaenoic acid.
\newblock Journal of Biological Chemistry. 1998;273(18):11056--11061.

\bibitem{meves2008ara}
Meves H.
\newblock Arachidonic acid and ion channels: an update.
\newblock British journal of pharmacology. 2008;155(1):4--16.

\bibitem{matta2007trp}
Matta JA, Miyares RL, Ahern GP.
\newblock {TRPV1} is a novel target for omega-3 polyunsaturated fatty acids.
\newblock The Journal of physiology. 2007;578(2):397--411.

\bibitem{schweitzer1990ara}
Schweitzer P, Madamba S, Siggins GR.
\newblock Arachidonic acid metabolites as mediators of somatostatin-induced
  increase of neuronal {M}-current.
\newblock Nature. 1990;346(6283):464--467.

\bibitem{wang2021bas}
Wang F, Trier AM, Li F, Kim S, Chen Z, Chai JN, et~al.
\newblock A basophil-neuronal axis promotes itch.
\newblock Cell. 2021;184(2):422--440.

\bibitem{fernandes2013sup}
Fernandes ES, Vong CT, Quek S, Cheong J, Awal S, Gentry C, et~al.
\newblock Superoxide generation and leukocyte accumulation: key elements in the
  mediation of leukotriene {B}4-induced itch by transient receptor potential
  ankyrin 1 and transient receptor potential vanilloid 1.
\newblock The FASEB Journal. 2013;27(4):1664--1673.

\bibitem{bazan2011end}
Bazan NG, Musto AE, Knott EJ.
\newblock Endogenous signaling by omega-3 docosahexaenoic acid-derived
  mediators sustains homeostatic synaptic and circuitry integrity.
\newblock Molecular neurobiology. 2011;44:216--222.

\bibitem{khasabova2020int}
Khasabova IA, Golovko MY, Golovko SA, Simone DA, Khasabov SG.
\newblock Intrathecal administration of Resolvin D1 and E1 decreases
  hyperalgesia in mice with bone cancer pain: Involvement of endocannabinoid
  signaling.
\newblock Prostaglandins \& other lipid mediators. 2020;151:106479.

\bibitem{payrits2020res}
Payrits M, Horv{\'a}th {\'A}, Bir{\'o}-S{\"u}t{\H{o}} T, Erosty{\'a}k J, Makkai
  G, S{\'a}ghy {\'E}, et~al.
\newblock Resolvin {D1} and {D2} inhibit transient receptor potential vanilloid
  1 and ankyrin 1 ion channel activation on sensory neurons via lipid raft
  modification.
\newblock International Journal of Molecular Sciences. 2020;21(14):5019.

\bibitem{cristino2014ECB}
Cristino L, Becker T, di~Marzo V.
\newblock Endocannabinoids and energy homeostasis: an update.
\newblock Biofactors. 2014;40(4):389--397.

\bibitem{fujita2001doc}
Fujita S, Ikegaya Y, Nishikawa M, Nishiyama N, Matsuki N.
\newblock Docosahexaenoic acid improves long-term potentiation attenuated by
  phospholipase {A}2 inhibitor in rat hippocampal slices.
\newblock British journal of pharmacology. 2001;132(7):1417.

\bibitem{barbosa2022mic}
Barbosa-Silva MC, Campos RMP, Del~Castilo I, Fran{\c{c}}a JV, Frost PS, Penido
  C, et~al.
\newblock Mice lacking 5-lipoxygenase display motor deficits associated with
  cortical and hippocampal synapse abnormalities.
\newblock Brain, Behavior, and Immunity. 2022;100:183--193.

\bibitem{kim2011syn}
Kim HY, Spector AA, Xiong ZM.
\newblock A synaptogenic amide {N}-docosahexaenoylethanolamide promotes
  hippocampal development.
\newblock Prostaglandins \& other lipid mediators. 2011;96(1-4):114--120.

\bibitem{wang2015res}
Wang X, Zhu M, Hjorth E, Cort{\'e}s-Toro V, Eyjolfsdottir H, Graff C, et~al.
\newblock Resolution of inflammation is altered in {A}lzheimer's disease.
\newblock Alzheimer's \& Dementia. 2015;11(1):40--50.

\bibitem{shalini2018dis}
Shalini SM, Ho CFY, Ng YK, Tong JX, Ong ES, Herr DR, et~al.
\newblock Distribution of {A}lox15 in the rat brain and its role in prefrontal
  cortical resolvin {D1} formation and spatial working memory.
\newblock Molecular Neurobiology. 2018;55:1537--1550.

\bibitem{gray2015CRH}
Gray JM, Vecchiarelli HA, Morena M, Lee TT, Hermanson DJ, Kim AB, et~al.
\newblock Corticotropin-releasing hormone drives anandamide hydrolysis in the
  amygdala to promote anxiety.
\newblock J Neurosci. 2015;35(9):3879--3892.

\bibitem{gray2014ECB}
Gray JM, Vecchiarelli HA, Hill MN.
\newblock Endocannabinoid signaling and synaptic plasticity during stress.
\newblock In: Synaptic Stress and Pathogenesis of Neuropsychiatric Disorders.
  Springer; 2014. p. 99--124.

\bibitem{balsevich2017ECB}
Balsevich G, Petrie GN, Hill MN.
\newblock Endocannabinoids: Effectors of glucocorticoid signaling.
\newblock Frontiers in Neuroendocrinology. 2017;47:86--108.

\bibitem{chameau2007glu}
Chameau P, Qin Y, Spijker S, Smit G, Jo{\"e}ls M.
\newblock Glucocorticoids specifically enhance {L}-type calcium current
  amplitude and affect calcium channel subunit expression in the mouse
  hippocampus.
\newblock Journal of neurophysiology. 2007;97(1):5--14.

\bibitem{kuster2010redox}
Kuster GM, Lancel S, Zhang J, Communal C, Trucillo MP, Lim CC, et~al.
\newblock Redox-mediated reciprocal regulation of {SERCA} and {Na+}--{Ca2+}
  exchanger contributes to sarcoplasmic reticulum {Ca2+} depletion in cardiac
  myocytes.
\newblock Free Radical Biology and Medicine. 2010;48(9):1182--1187.

\bibitem{wei2009pre}
Wei Y, Wang X, Wang L.
\newblock Presence and regulation of cannabinoid receptors in human retinal
  pigment epithelial cells.
\newblock Molecular vision. 2009;15:1243.

\bibitem{sandman2016neu}
Sandman CA, Glynn LM, Davis EP.
\newblock Neurobehavioral consequences of fetal exposure to gestational stress.
\newblock Fetal development: Research on brain and behavior, environmental
  influences, and emerging technologies. 2016;p. 229--265.

\bibitem{meyer2019neu}
Meyer U.
\newblock Neurodevelopmental resilience and susceptibility to maternal immune
  activation.
\newblock Trends in neurosciences. 2019;42(11):793--806.

\bibitem{babenko2015str}
Babenko O, Kovalchuk I, Metz GA.
\newblock Stress-induced perinatal and transgenerational epigenetic programming
  of brain development and mental health.
\newblock Neuroscience \& Biobehavioral Reviews. 2015;48:70--91.

\bibitem{short2019ear}
Short AK, Baram TZ.
\newblock Early-life adversity and neurological disease: age-old questions and
  novel answers.
\newblock Nature Reviews Neurology. 2019;15(11):657--669.

\bibitem{eiland2013str}
EILAND L, ROMEO R.
\newblock Stress and the developing adolescent brain.
\newblock Neuroscience. 2013;249:162--171.

\bibitem{szyf2015non}
Szyf M.
\newblock Nongenetic inheritance and transgenerational epigenetics.
\newblock Trends in molecular medicine. 2015;21(2):134--144.

\bibitem{dick2021str}
Dick ALW, Chen A.
\newblock Stress-Mediated Regulation of the {DNA} Methylome.
\newblock In: Stress: Genetics, Epigenetics and Genomics. Elsevier; 2021. p.
  37--47.

\bibitem{goldstein2021ear}
Goldstein~Ferber S, Trezza V, Weller A.
\newblock Early life stress and development of the endocannabinoid system: A
  bidirectional process in programming future coping.
\newblock Developmental Psychobiology. 2021;63(2):143--152.

\bibitem{rusconi2020end}
Rusconi F, Rubino T, Battaglioli E.
\newblock Endocannabinoid-epigenetic cross-talk: a bridge toward stress coping.
\newblock International Journal of Molecular Sciences. 2020;21(17):6252.

\bibitem{demaili2023epi}
Demaili A, Portugalov A, Dudai M, Maroun M, Akirav I, Braun K, et~al.
\newblock Epigenetic (re) programming of gene expression changes of {CB1R} and
  {FAAH} in the medial prefrontal cortex in response to early life and
  adolescence stress exposure.
\newblock Frontiers in Cellular Neuroscience. 2023;17:1129946.

\bibitem{zumbrun2015epi}
Zumbrun EE, Sido JM, Nagarkatti PS, Nagarkatti M.
\newblock Epigenetic regulation of immunological alterations following prenatal
  exposure to marijuana cannabinoids and its long term consequences in
  offspring.
\newblock Journal of Neuroimmune Pharmacology. 2015;10:245--254.

\bibitem{franklin2010epi}
Franklin TB, Russig H, Weiss IC, Gr{\"a}ff J, Linder N, Michalon A, et~al.
\newblock Epigenetic transmission of the impact of early stress across
  generations.
\newblock Biological psychiatry. 2010;68(5):408--415.

\bibitem{hong2015epi}
Hong S, Zheng G, Wiley JW.
\newblock Epigenetic regulation of genes that modulate chronic stress-induced
  visceral pain in the peripheral nervous system.
\newblock Gastroenterology. 2015;148(1):148--157.

\bibitem{wiedmann2022dna}
Wiedmann M, Kuitunen-Paul S, Basedow LA, Wolff M, DiDonato N, Franzen J, et~al.
\newblock {DNA} methylation changes associated with cannabis use and verbal
  learning performance in adolescents: an exploratory whole genome methylation
  study.
\newblock Translational Psychiatry. 2022;12(1):317.

\bibitem{hapgood2016glu}
Hapgood JP, Avenant C, Moliki JM.
\newblock Glucocorticoid-independent modulation of {GR} activity: Implications
  for immunotherapy.
\newblock Pharmacology \& therapeutics. 2016;165:93--113.

\bibitem{hendrik2020con}
Hendrik~Schmidt J, Perslev M, Bukowski L, Stoklund M, Herborg F, Herlo R,
  et~al.
\newblock Constitutive internalization across therapeutically targeted {GPCR}s
  correlates with constitutive activity.
\newblock Basic \& Clinical Pharmacology \& Toxicology. 2020;126:116--121.

\bibitem{robinson2010eff}
Robinson SA, Loiacono RE, Christopoulos A, Sexton PM, Malone DT.
\newblock The effect of social isolation on rat brain expression of genes
  associated with endocannabinoid signaling.
\newblock Brain research. 2010;1343:153--167.

\bibitem{sciolino2010soc}
SCIOLINO N, BORTOLATO M, EISENSTEIN S, FU J, OVEISI F, HOHMANN A.
\newblock Social isolation and chronic handling alter endocannabinoid signaling
  and behavioral reactivity to context in adult rats.
\newblock Neuroscience. 2010;168(2):371--386.

\bibitem{zamberletti2012chr}
Zamberletti E, Piscitelli F, Rubino T, Di~Marzo V, Parolaro D.
\newblock Chronic blockade of {CB1} receptors reverses startle gating deficits
  and associated neurochemical alterations in rats reared in isolation.
\newblock British journal of pharmacology. 2012;167(8):1652--1664.

\bibitem{marco2014con}
Marco EM, Echeverry-Alzate V, L{\'o}pez-Moreno JA, Gin{\'e} E, Pe{\~n}asco S,
  Viveros MP.
\newblock Consequences of early life stress on the expression of
  endocannabinoid-related genes in the rat brain.
\newblock Behavioural pharmacology. 2014;25(5 and 6):547--556.

\bibitem{schneider2016adv}
Schneider P, Bindila L, Schmahl C, Bohus M, Meyer-Lindenberg A, Lutz B, et~al.
\newblock Adverse social experiences in adolescent rats result in enduring
  effects on social competence, pain sensitivity and endocannabinoid signaling.
\newblock Frontiers in behavioral neuroscience. 2016;10:203.

\bibitem{boero2018imp}
Boero G, Pisu MG, Biggio F, Muredda L, Carta G, Banni S, et~al.
\newblock Impaired glucocorticoid-mediated {HPA} axis negative feedback induced
  by juvenile social isolation in male rats.
\newblock Neuropharmacology. 2018;133:242--253.

\bibitem{uz2001glu}
Uz T, Dwivedi Y, Qeli A, Peters-Golden M, Pandey G, Manev H.
\newblock Glucocorticoid receptors are required for up-regulation of neuronal
  5-lipoxygenase (5{LOX}) expression by dexamethasone.
\newblock The FASEB Journal. 2001;15(10):1792--1794.

\bibitem{elliott2017rol}
Elliott E, Hanson C, Anderson-Berry A, Nordgren T.
\newblock The role of specialized pro-resolving mediators in maternal-fetal
  health.
\newblock Prostaglandins, Leukotrienes and Essential Fatty Acids.
  2017;126:98--104.

\bibitem{imbesi2009lox}
Imbesi M, Dzitoyeva S, Ng LW, Manev H.
\newblock 5-Lipoxygenase and epigenetic {DNA} methylation in aging cultures of
  cerebellar granule cells.
\newblock Neuroscience. 2009;164(4):1531--1537.

\bibitem{gattaz1987inc}
Gattaz WF, K{\"o}llisch M, Thuren T, Virtanen JA, Kinnunen PK.
\newblock Increased plasma phospholipase-{A}2 activity in schizophrenic
  patients: reduction after neuroleptic therapy.
\newblock Biological Psychiatry. 1987;22(4):421--426.

\bibitem{noponen1993ele}
Noponen M, Sanfilipo M, Samanich K, Ryer H, Ko G, Angrist B, et~al.
\newblock Elevated {PLA2} activity in schizophrenics and other psychiatric
  patients.
\newblock Biological psychiatry. 1993;34(9):641--649.

\bibitem{gattaz1995inc}
Gattaz WF, Schmitt A, Maras A.
\newblock Increased platelet phospholipase {A}2 activity in schizophrenia.
\newblock Schizophrenia research. 1995;16(1):1--6.

\bibitem{ross1997inc}
Ross BM, Hudson C, Erlich J, Warsh JJ, Kish SJ.
\newblock Increased phospholipid breakdown in schizophrenia: evidence for the
  involvement of a calcium-independent phospholipase {A}2.
\newblock Archives of general psychiatry. 1997;54(5):487--494.

\bibitem{ross1999dif}
Ross BM, Turenne S, Moszczynska A, Warsh JJ, Kish SJ.
\newblock Differential alteration of phospholipase {A}2 activities in brain of
  patients with schizophrenia.
\newblock Brain research. 1999;821(2):407--413.

\bibitem{tavares2003inc}
Tavares~Jr H, Yacubian J, Talib LL, Barbosa NR, Gattaz WF.
\newblock Increased phospholipase {A2} activity in schizophrenia with absent
  response to niacin.
\newblock Schizophrenia research. 2003;61(1):1--6.

\bibitem{smesny2005inc}
Smesny S, Kinder D, Willhardt I, Rosburg T, Lasch J, Berger G, et~al.
\newblock Increased calcium-independent phospholipase {A2} activity in first
  but not in multiepisode chronic schizophrenia.
\newblock Biological psychiatry. 2005;57(4):399--405.

\bibitem{smesny2010pho}
Smesny S, Milleit B, Nenadic I, Preul C, Kinder D, Lasch J, et~al.
\newblock Phospholipase {A2} activity is associated with structural brain
  changes in schizophrenia.
\newblock Neuroimage. 2010;52(4):1314--1327.

\bibitem{smesny2011pho}
Smesny S, Kunstmann C, Kunstmann S, Willhardt I, Lasch J, Yotter RA, et~al.
\newblock Phospholipase {A2} activity in first episode schizophrenia:
  associations with symptom severity and outcome at week 12.
\newblock The World Journal of Biological Psychiatry. 2011;12(8):598--607.

\bibitem{xu2019inv}
Xu C, Yang X, Sun L, Yang T, Cai C, Wang P, et~al.
\newblock An investigation of calcium-independent phospholipase {A}2 (i{PLA2})
  and cytosolic phospholipase {A}2 (c{PLA}2) in schizophrenia.
\newblock Psychiatry Research. 2019;273:782--787.

\bibitem{yang2021dys}
Yang X, Li M, Jiang J, Hu X, Qing Y, Sun L, et~al.
\newblock Dysregulation of phospholipase and cyclooxygenase expression is
  involved in Schizophrenia.
\newblock EBioMedicine. 2021;64:103239.

\bibitem{talib2021inc}
Talib LL, Costa AC, Joaquim HP, Pereira CA, Van~de Bilt MT, Loch AA, et~al.
\newblock Increased {PLA} 2 activity in individuals at ultra-high risk for
  psychosis.
\newblock European archives of psychiatry and clinical neuroscience;p. 1--7.

\bibitem{li2022imp}
Li M, Gao Y, Wang D, Hu X, Jiang J, Qing Y, et~al.
\newblock Impaired Membrane Lipid Homeostasis in Schizophrenia.
\newblock Schizophrenia Bulletin. 2022;48(5):1125--1135.

\bibitem{albers1993pho}
Albers M, Meurer H, M{\"a}rki F, Klotz J.
\newblock Phospholipase {A}2 activity in serum of neuroleptic-naive psychiatric
  inpatients.
\newblock Pharmacopsychiatry. 1993;26(03):94--98.

\bibitem{clancy2014pre}
Clancy MJ, Clarke MC, Connor DJ, Cannon M, Cotter DR.
\newblock The prevalence of psychosis in epilepsy; a systematic review and
  meta-analysis.
\newblock BMC psychiatry. 2014;14(1):1--9.

\bibitem{gattaz2011inc}
Gattaz WF, Valente KD, Raposo NR, Vincentiis S, Talib LL.
\newblock Increased {PLA2} activity in the hippocampus of patients with
  temporal lobe epilepsy and psychosis.
\newblock Journal of psychiatric research. 2011;45(12):1617--1620.

\bibitem{elliott2009del}
Elliott B, Joyce E, Shorvon S.
\newblock Delusions, illusions and hallucinations in epilepsy: 2. Complex
  phenomena and psychosis.
\newblock Epilepsy research. 2009;85(2-3):172--186.

\bibitem{onder2004nsa}
Onder G, Pellicciotti F, Gambassi G, Bernabei R.
\newblock {NSAID}-related psychiatric adverse events: who is at risk?
\newblock Drugs. 2004;64:2619--2627.

\bibitem{lu2001dif}
Lu XR, Ong WY, Halliwell B, Horrocks LA, Farooqui AA.
\newblock Differential effects of calcium-dependent and calcium-independent
  phospholipase {A2} inhibitors on kainate-induced neuronal injury in rat
  hippocampal slices.
\newblock Free Radical Biology and Medicine. 2001;30(11):1263--1273.

\bibitem{charlson2018glo}
Charlson FJ, Ferrari AJ, Santomauro DF, Diminic S, Stockings E, Scott JG,
  et~al.
\newblock Global epidemiology and burden of schizophrenia: findings from the
  global burden of disease study 2016.
\newblock Schizophrenia bulletin. 2018;44(6):1195--1203.

\bibitem{peet1995dep}
Peet M, Laugharne J, Rangarajan N, Horrobin D, Reynolds G.
\newblock Depleted red cell membrane essential fatty acids in drug-treated
  schizophrenic patients.
\newblock Journal of psychiatric research. 1995;29(3):227--232.

\bibitem{yao1996abn}
Yao JK, van Kammen DP, Gurklis JA.
\newblock Abnormal incorporation of arachidonic acid into platelets of
  drug-free patients with schizophrenia.
\newblock Psychiatry research. 1996;60(1):11--21.

\bibitem{yao2000mem}
Yao JK, Leonard S, Reddy RD.
\newblock Membrane phospholipid abnormalities in postmortem brains from
  schizophrenic patients.
\newblock Schizophrenia research. 2000;42(1):7--17.

\bibitem{khan2002red}
Khan MM, Evans DR, Gunna V, Scheffer RE, Parikh VV, Mahadik SP.
\newblock Reduced erythrocyte membrane essential fatty acids and increased
  lipid peroxides in schizophrenia at the never-medicated first-episode of
  psychosis and after years of treatment with antipsychotics.
\newblock Schizophrenia research. 2002;58(1):1--10.

\bibitem{arvindakshan2003ess}
Arvindakshan M, Sitasawad S, Debsikdar V, Ghate M, Evans D, Horrobin DF, et~al.
\newblock Essential polyunsaturated fatty acid and lipid peroxide levels in
  never-medicated and medicated schizophrenia patients.
\newblock Biological psychiatry. 2003;53(1):56--64.

\bibitem{reddy2004red}
Reddy RD, Keshavan MS, Yao JK.
\newblock Reduced red blood cell membrane essential polyunsaturated fatty acids
  in first episode schizophrenia at neuroleptic-naive baseline.
\newblock Schizophrenia bulletin. 2004;30(4):901--911.

\bibitem{kemperman2006low}
Kemperman R, Veurink M, van~der Wal T, Knegtering H, Bruggeman R, Fokkema M,
  et~al.
\newblock Low essential fatty acid and {B}-vitamin status in a subgroup of
  patients with schizophrenia and its response to dietary supplementation.
\newblock Prostaglandins, leukotrienes and essential fatty acids.
  2006;74(2):75--85.

\bibitem{mcnamara2007abn}
McNamara RK, Jandacek R, Rider T, Tso P, Hahn CG, Richtand NM, et~al.
\newblock Abnormalities in the fatty acid composition of the postmortem
  orbitofrontal cortex of schizophrenic patients: gender differences and
  partial normalization with antipsychotic medications.
\newblock Schizophrenia research. 2007;91(1-3):37--50.

\bibitem{sethom2010pol}
Sethom M, Fares S, Bouaziz N, Melki W, Jemaa R, Feki M, et~al.
\newblock Polyunsaturated fatty acids deficits are associated with psychotic
  state and negative symptoms in patients with schizophrenia.
\newblock Prostaglandins, leukotrienes and essential fatty acids.
  2010;83(3):131--136.

\bibitem{rice2015ery}
Rice SM, Sch{\"a}fer MR, Klier C, Mossaheb N, Vijayakumar N, Amminger GP.
\newblock Erythrocyte polyunsaturated fatty acid levels in young people at
  ultra-high risk for psychotic disorder and healthy adolescent controls.
\newblock Psychiatry research. 2015;228(1):174--176.

\bibitem{hamazaki2016fat}
Hamazaki K, Maekawa M, Toyota T, Iwayama Y, Dean B, Hamazaki T, et~al.
\newblock Fatty acid composition and fatty acid binding protein expression in
  the postmortem frontal cortex of patients with schizophrenia: A case--control
  study.
\newblock Schizophrenia research. 2016;171(1-3):225--232.

\bibitem{wang2018alt}
Wang D, Sun X, Yan J, Ren B, Cao B, Lu Q, et~al.
\newblock Alterations of eicosanoids and related mediators in patients with
  schizophrenia.
\newblock Journal of psychiatric research. 2018;102:168--178.

\bibitem{alqarni2020com}
Alqarni A, Mitchell TW, McGorry PD, Nelson B, Markulev C, Yuen HP, et~al.
\newblock Comparison of erythrocyte omega-3 index, fatty acids and molecular
  phospholipid species in people at ultra-high risk of developing psychosis and
  healthy people.
\newblock Schizophrenia research. 2020;226:44--51.

\bibitem{zhou2020red}
Zhou X, Long T, Haas GL, Cai H, Yao JK.
\newblock Reduced levels and disrupted biosynthesis pathways of plasma free
  fatty acids in first-episode antipsychotic-na{\"\i}ve schizophrenia patients.
\newblock Frontiers in Neuroscience. 2020;14:784.

\bibitem{li2022red}
Li N, Yang P, Tang M, Liu Y, Guo W, Lang B, et~al.
\newblock Reduced erythrocyte membrane polyunsaturated fatty acid levels
  indicate diminished treatment response in patients with multi-versus
  first-episode schizophrenia.
\newblock Schizophrenia. 2022;8(1):7.

\bibitem{yu2022nia}
Yu YH, Su HM, Lin SH, Hsiao PC, Lin YT, Liu CM, et~al.
\newblock Niacin skin flush and membrane polyunsaturated fatty acids in
  schizophrenia from the acute state to partial remission: a dynamic
  relationship.
\newblock Schizophrenia. 2022;8(1):38.

\bibitem{das1998inc}
Das I, Khan N.
\newblock Increased arachidonic acid induced platelet chemiluminescence
  indicates cyclooxygenase overactivity in schizophrenic subjects.
\newblock Prostaglandins, leukotrienes and essential fatty acids.
  1998;58(3):165--168.

\bibitem{berger2019rel}
Berger M, Nelson B, Markulev C, Yuen HP, Sch{\"a}fer MR, Mossaheb N, et~al.
\newblock Relationship between polyunsaturated fatty acids and psychopathology
  in the {NEURAPRO} clinical trial.
\newblock Frontiers in psychiatry. 2019;10:393.

\bibitem{kale2008opp}
Kale A, Joshi S, Naphade N, Sapkale S, Raju M, Pillai A, et~al.
\newblock Opposite changes in predominantly docosahexaenoic acid ({DHA}) in
  cerebrospinal fluid and red blood cells from never-medicated first-episode
  psychotic patients.
\newblock Schizophrenia Research. 2008;98(1-3):295--301.

\bibitem{mahadik1996pla}
Mahadik SP, Mukherjee S, Horrobin DF, Jenkins K, Correnti EE, Scheffer RE.
\newblock Plasma membrane phospholipid fatty acid composition of cultured skin
  fibroblasts from schizophrenic patients: comparison with bipolar patients and
  normal subjects.
\newblock Psychiatry Research. 1996;63(2-3):133--142.

\bibitem{frajerman2023mem}
Frajerman A, Chaumette B, Farabos D, Despres G, Simonard C, Lamazi{\`e}re A,
  et~al.
\newblock Membrane Lipids in Ultra-High-Risk Patients: Potential Predictive
  Biomarkers of Conversion to Psychosis.
\newblock Nutrients. 2023;15(9):2215.

\bibitem{smesny2014ome}
Smesny S, Milleit B, Hipler U, Milleit C, Sch{\"a}fer M, Klier C, et~al.
\newblock Omega-3 fatty acid supplementation changes intracellular
  phospholipase {A2} activity and membrane fatty acid profiles in individuals
  at ultra-high risk for psychosis.
\newblock Molecular psychiatry. 2014;19(3):317--324.

\bibitem{medema2016lev}
Medema S, Mocking RJ, Koeter MW, Vaz FM, Meijer C, de~Haan L, et~al.
\newblock Levels of red blood cell fatty acids in patients with psychosis,
  their unaffected siblings, and healthy controls.
\newblock Schizophrenia bulletin. 2016;42(2):358--368.

\bibitem{watari2010hos}
Watari M, Hamazaki K, Hirata T, Hamazaki T, Okubo Y.
\newblock Hostility of drug-free patients with schizophrenia and n-3
  polyunsaturated fatty acid levels in red blood cells.
\newblock Psychiatry Research. 2010;177(1-2):22--26.

\bibitem{szeszko2021lon}
Szeszko PR, McNamara RK, Gallego JA, Malhotra AK, Govindarajulu U, Peters BD,
  et~al.
\newblock Longitudinal investigation of the relationship between omega-3
  polyunsaturated fatty acids and neuropsychological functioning in
  recent-onset psychosis: a randomized clinical trial.
\newblock Schizophrenia research. 2021;228:180--187.

\bibitem{maida2006cyt}
Maida ME, Hurley SD, Daeschner JA, Moore AH, O'banion MK.
\newblock Cytosolic prostaglandin {E2} synthase (c{PGES}) expression is
  decreased in discrete cortical regions in psychiatric disease.
\newblock Brain research. 2006;1103(1):164--172.

\bibitem{dean2018stu}
Dean B, Gibbons A, Gogos A, Udawela M, Thomas E, Scarr E.
\newblock Studies on prostaglandin-endoperoxide synthase 1: lower levels in
  schizophrenia and after treatment with antipsychotic drugs in conjunction
  with aspirin.
\newblock International Journal of Neuropsychopharmacology.
  2018;21(3):216--225.

\bibitem{ganzinelli2010aut}
Ganzinelli S, Borda E, Sterin-Borda L.
\newblock Autoantibodies from schizophrenia patients induce cerebral
  {COX}-1/i{NOS} m{RNA} expression with {NO/PGE2/MMP}-3 production.
\newblock International Journal of Neuropsychopharmacology.
  2010;13(3):293--303.

\bibitem{abdulla1975eff}
Abdulla Y, Hamadah K.
\newblock Effect of {ADP} on {PGE1} formation in blood platelets from patients
  with depression, mania and schizophrenia.
\newblock The British Journal of Psychiatry. 1975;127(6):591--595.

\bibitem{kaiya1991pro}
Kaiya H.
\newblock Prostaglandin {E1} suppression of platelet aggregation response in
  schizophrenia.
\newblock Schizophrenia research. 1991;5(1):67--80.

\bibitem{tang2012dif}
Tang B, Capitao C, Dean B, Thomas EA.
\newblock Differential age-and disease-related effects on the expression of
  genes related to the arachidonic acid signaling pathway in schizophrenia.
\newblock Psychiatry research. 2012;196(2-3):201--206.

\bibitem{garcia2018reg}
Caso J, Garc{\'\i}a-Portilla M, De~la Fuente-Tom{\'a}s L, Gonz{\'a}lez-Blanco
  L, Mart{\'\i}nez PS, Leza J, et~al.
\newblock Regulation of inflammatory pathways in schizophrenia: a comparative
  study with bipolar disorder and healthy controls.
\newblock European Psychiatry. 2018;47:50--59.

\bibitem{yuksel2019ser}
Y{\"u}ksel RN, Titiz AP, Yaylac{\i} ET, {\"U}nal K, Turhan T, Erzin G, et~al.
\newblock Serum {PGE2}, 15d-{PGJ, PPAR}$\gamma$ and {CRP} levels in patients
  with schizophrenia.
\newblock Asian journal of psychiatry. 2019;46:24--28.

\bibitem{fromer2016gen}
Fromer M, Roussos P, Sieberts SK, Johnson JS, Kavanagh DH, Perumal TM, et~al.
\newblock Gene expression elucidates functional impact of polygenic risk for
  schizophrenia.
\newblock Nature neuroscience. 2016;19(11):1442--1453.

\bibitem{kaiya1989ele}
Kaiya H, Uematsu M, Ofuji M, Nishida A, Takeuchi K, Nozaki M, et~al.
\newblock Elevated plasma prostaglandin {E2} levels in schizophrenia.
\newblock Journal of neural transmission. 1989;77:39--46.

\bibitem{martinez2011ant}
Mart{\'\i}nez-Gras I, P{\'e}rez-Nievas BG, Garc{\'\i}a-Bueno B, Madrigal JL,
  Andr{\'e}s-Esteban E, Rodr{\'\i}guez-Jim{\'e}nez R, et~al.
\newblock The anti-inflammatory prostaglandin 15d-{PGJ2} and its nuclear
  receptor {PPAR}gamma are decreased in schizophrenia.
\newblock Schizophrenia research. 2011;128(1-3):15--22.

\bibitem{garcia2014pro}
Garc{\'\i}a-Bueno B, Bioque M, Mac-Dowell KS, Barcones MF,
  Mart{\'\i}nez-Cengotitabengoa M, Pina-Camacho L, et~al.
\newblock Pro-/anti-inflammatory dysregulation in patients with first episode
  of psychosis: toward an integrative inflammatory hypothesis of schizophrenia.
\newblock Schizophrenia bulletin. 2014;40(2):376--387.

\bibitem{cabrera2016cog}
Cabrera B, Penad{\'e}s R, Gonz{\'a}lez-Pinto A, Parellada M, Bobes J, Lobo A,
  et~al.
\newblock Cognition and psychopathology in first-episode psychosis: are they
  related to inflammation?
\newblock Psychological medicine. 2016;46(10):2133--2144.

\bibitem{pereira2022cox}
Pereira CA, Costa AC, Joaquim HP, Talib LL, van~de Bilt MT, Loch AA, et~al.
\newblock {COX-2} pathway is upregulated in ultra-high risk individuals for
  psychosis.
\newblock The World Journal of Biological Psychiatry. 2022;23(3):236--241.

\bibitem{nadalin2010nia}
Nadalin S, Bureti{\'c}-Tomljanovi{\'c} A, Rube{\v{s}}a G, Tomljanovi{\'c} D,
  Gudelj L.
\newblock Niacin skin flush test: a research tool for studying schizophrenia.
\newblock Psychiatria Danubina. 2010;22(1):14--27.

\bibitem{dong2016dif}
Dong L, Zou H, Yuan C, Hong YH, Kuklev DV, Smith WL.
\newblock Different fatty acids compete with arachidonic acid for binding to
  the allosteric or catalytic subunits of cyclooxygenases to regulate
  prostanoid synthesis.
\newblock Journal of Biological Chemistry. 2016;291(8):4069--4078.

\bibitem{lacombe2018bra}
Lacombe RS, Chouinard-Watkins R, Bazinet RP.
\newblock Brain docosahexaenoic acid uptake and metabolism.
\newblock Molecular Aspects of Medicine. 2018;64:109--134.

\bibitem{mcnamara2013adu}
McNamara RK, Jandacek R, Rider T, Tso P, Dwivedi Y, Pandey GN.
\newblock Adult medication-free schizophrenic patients exhibit long-chain
  omega-3 fatty acid deficiency: implications for cardiovascular disease risk.
\newblock Cardiovascular psychiatry and neurology. 2013;2013.

\bibitem{mcevoy2013lip}
McEvoy J, Baillie RA, Zhu H, Buckley P, Keshavan MS, Nasrallah HA, et~al.
\newblock Lipidomics reveals early metabolic changes in subjects with
  schizophrenia: effects of atypical antipsychotics.
\newblock PloS one. 2013;8(7):e68717.

\bibitem{assies2001sig}
Assies J, Lieverse R, Vreken P, Wanders RJ, Dingemans PM, Linszen DH.
\newblock Significantly reduced docosahexaenoic and docosapentaenoic acid
  concentrations in erythrocyte membranes from schizophrenic patients compared
  with a carefully matched control group.
\newblock Biological psychiatry. 2001;49(6):510--522.

\bibitem{wood2015dys}
Wood PL, Unfried G, Whitehead W, Phillipps A, Wood JA.
\newblock Dysfunctional plasmalogen dynamics in the plasma and platelets of
  patients with schizophrenia.
\newblock Schizophrenia Research. 2015;161(2-3):506--510.

\bibitem{mahadik1996uti}
Mahadik S, Scheffer R, Mukherjee S, Correnti E.
\newblock Utilization of precursor essential fatty acids in culture by skin
  fibroblasts from schizophrenic patients and normal controls.
\newblock Prostaglandins, leukotrienes and essential fatty acids.
  1996;55(1-2):65--70.

\bibitem{berger2017om}
Berger M, Smesny S, Kim S, Davey C, Rice S, Sarnyai Z, et~al.
\newblock Omega-6 to omega-3 polyunsaturated fatty acid ratio and subsequent
  mood disorders in young people with at-risk mental states: a 7-year
  longitudinal study.
\newblock Translational psychiatry. 2017;7(8):e1220--e1220.

\bibitem{pettegrew1991alt}
Pettegrew JW, Keshavan MS, Panchalingam K, Strychor S, Kaplan DB, Tretta MG,
  et~al.
\newblock Alterations in brain high-energy phosphate and membrane phospholipid
  metabolism in first-episode, drug-naive schizophrenics: A pilot study of the
  dorsal prefrontal cortex by in vivo phosphorus 31 nuclear magnetic resonance
  spectroscopy.
\newblock Archives of general psychiatry. 1991;48(6):563--568.

\bibitem{keshavan1993ery}
Keshavan MS, Mallinger AG, Pettegrew JW, Dippold C.
\newblock Erythrocyte membrane phospholipids in psychotic patients.
\newblock Psychiatry Research. 1993;49(1):89--95.

\bibitem{mahadik1994pla}
Mahadik SP, Mukherjee S, Correnti EE, Kelkar HS, Wakade CG, Costa RM, et~al.
\newblock Plasma membrane phospholipid and cholesterol distribution of skin
  fibroblasts from drug-naive patients at the onset of psychosis.
\newblock Schizophrenia research. 1994;13(3):239--247.

\bibitem{stanley1995viv}
Stanley JA, Williamson PC, Drost DJ, Carr TJ, Rylett RJ, Malla A, et~al.
\newblock An in vivo study of the prefrontal cortex of schizophrenic patients
  at different stages of illness via phosphorus magnetic resonance
  spectroscopy.
\newblock Archives of general psychiatry. 1995;52(5):399--406.

\bibitem{schmitt2001eff}
Schmitt A, Maras A, Petroianu G, Braus D, Scheuer L.
\newblock Effects of antipsychotic treatment on membrane phospholipid
  metabolism in schizophrenia.
\newblock Journal of neural transmission. 2001;108:1081--1091.

\bibitem{ponizovsky2001pho}
Ponizovsky A, Modai I, Nechamkin Y, Barshtein G, Ritsner M, Yedgar S, et~al.
\newblock Phospholipid patterns of erythrocytes in schizophrenia: relationships
  to symptomatology.
\newblock Schizophrenia research. 2001;52(1-2):121--126.

\bibitem{ryazantseva2002cha}
Ryazantseva N, Novitsky V, Vlasova N, Tokareva N, Zhukova O, Mikheev S, et~al.
\newblock Changes in the Composition, Structure, and Metabolism of Erythrocyte
  Membranes in Schizophrenia and Other Psychopathologies.
\newblock Neurophysiology. 2002;34:332--341.

\bibitem{schmitt2004alt}
Schmitt A, Wilczek K, Blennow K, Maras A, Jatzko A, Petroianu G, et~al.
\newblock Altered thalamic membrane phospholipids in schizophrenia: a
  postmortem study.
\newblock Biological psychiatry. 2004;56(1):41--45.

\bibitem{kaddurah2012imp}
Kaddurah-Daouk R, McEvoy J, Baillie R, Zhu H, Yao JK, Nimgaonkar VL, et~al.
\newblock Impaired plasmalogens in patients with schizophrenia.
\newblock Psychiatry research. 2012;198(3):347--352.

\bibitem{wang2019met}
Wang D, Cheng SL, Fei Q, Gu H, Raftery D, Cao B, et~al.
\newblock Metabolic profiling identifies phospholipids as potential serum
  biomarkers for schizophrenia.
\newblock Psychiatry research. 2019;272:18--29.

\bibitem{wang2021cha}
Wang D, Sun X, Maziade M, Mao W, Zhang C, Wang J, et~al.
\newblock Characterising phospholipids and free fatty acids in patients with
  schizophrenia: A case-control study.
\newblock The World Journal of Biological Psychiatry. 2021;22(3):161--174.

\bibitem{song2023pot}
Song M, Liu Y, Zhou J, Shi H, Su X, Shao M, et~al.
\newblock Potential plasma biomarker panels identification for the diagnosis of
  first-episode schizophrenia and monitoring antipsychotic monotherapy with the
  use of metabolomics analyses.
\newblock Psychiatry Research. 2023;321:115070.

\bibitem{nuss2009abn}
Nuss P, Tessier C, Ferreri F, De~Hert M, Peuskens J, Trugnan G, et~al.
\newblock Abnormal transbilayer distribution of phospholipids in red blood cell
  membranes in schizophrenia.
\newblock Psychiatry Research. 2009;169(2):91--96.

\bibitem{atack1998cer}
Atack JR, Levine J, Belmaker RH.
\newblock Cerebrospinal fluid inositol monophosphatase: elevated activity in
  depression and neuroleptic-treated schizophrenia.
\newblock Biological psychiatry. 1998;44(6):433--437.

\bibitem{shimon1998ino}
Shimon H, Sobolev Y, Davidson M, Haroutunian V, Belmaker RH, Agam G.
\newblock Inositol levels are decreased in postmortem brain of schizophrenic
  patients.
\newblock Biological Psychiatry. 1998;44(6):428--432.

\bibitem{jope1998sel}
Jope RS, Song L, Grimes CA, Pacheco MA, Dilley GE, Li X, et~al.
\newblock Selective increases in phosphoinositide signaling activity and {G}
  protein levels in postmortem brain from subjects with schizophrenia or
  alcohol dependence.
\newblock Journal of neurochemistry. 1998;70(2):763--771.

\bibitem{essali1990pla}
Essali MA, Das I, de~Belleroche J, Hirsch SR.
\newblock The platelet polyphosphoinositide system in schizophrenia: the
  effects of neuroleptic treatment.
\newblock Biological psychiatry. 1990;28(6):475--487.

\bibitem{das1992ino}
Das I, Essali M, De~Belleroche J, Hirsch S.
\newblock Inositol phospholipid turnover in platelets of schizophrenic
  patients.
\newblock Prostaglandins, leukotrienes and essential fatty acids.
  1992;46(1):65--66.

\bibitem{yao1992inc}
Yao J, Yasaei P, Van~Kammen D.
\newblock Increased turnover of platelet phosphatidylinositol in schizophrenia.
\newblock Prostaglandins, leukotrienes and essential fatty acids.
  1992;46(1):39--46.

\bibitem{rybakowski1997inc}
Rybakowski J, Lehmann W.
\newblock Increased erythrocyte inositol monophosphatase activity in
  schizophrenia.
\newblock European Psychiatry. 1997;12(1):44--45.

\bibitem{kunii2021evi}
Kunii Y, Matsumoto J, Izumi R, Nagaoka A, Hino M, Shishido R, et~al.
\newblock Evidence for altered phosphoinositide signaling-associated molecules
  in the postmortem prefrontal cortex of patients with schizophrenia.
\newblock International journal of molecular sciences. 2021;22(15):8280.

\bibitem{durrenberger2015com}
Durrenberger PF, Fernando FS, Kashefi SN, Bonnert TP, Seilhean D, Nait-Oumesmar
  B, et~al.
\newblock Common mechanisms in neurodegeneration and neuroinflammation: a
  {B}rain{N}et {E}urope gene expression microarray study.
\newblock Journal of neural transmission. 2015;122:1055--1068.

\bibitem{piras2019per}
Piras IS, Manchia M, Huentelman MJ, Pinna F, Zai CC, Kennedy JL, et~al.
\newblock Peripheral biomarkers in schizophrenia: a meta-analysis of microarray
  gene expression datasets.
\newblock International Journal of Neuropsychopharmacology.
  2019;22(3):186--193.

\bibitem{kim2010ass}
Kim T, Kim HJ, Park JK, Kim JW, Chung JH.
\newblock Association between polymorphisms of arachidonate 12-lipoxygenase
  ({ALOX12}) and schizophrenia in a Korean population.
\newblock Behavioral and Brain Functions. 2010;6(1):1--4.

\bibitem{akiyama2022chr}
Akiyama S, Nagai H, Oike S, Horikawa I, Shinohara M, Lu Y, et~al.
\newblock Chronic social defeat stress increases the amounts of 12-lipoxygenase
  lipid metabolites in the nucleus accumbens of stress-resilient mice.
\newblock Scientific Reports. 2022;12(1):11385.

\bibitem{junqueira2004all}
Junqueira R, Cordeiro Q, Meira-Lima I, Gattaz WF, Vallada H.
\newblock Allelic association analysis of phospholipase {A}2 genes with
  schizophrenia.
\newblock Psychiatric Genetics. 2004;14(3):157--160.

\bibitem{yu2005gen}
Yu Y, Tao R, Shi J, Zhang X, Kou C, Guo Y, et~al.
\newblock A genetic study of two calcium-independent cytosolic {PLA}2 genes in
  schizophrenia.
\newblock Prostaglandins, leukotrienes and essential fatty acids.
  2005;73(5):351--354.

\bibitem{tao2005cyt}
Tao R, Yu Y, Zhang X, Guo Y, Shi J, Zhang X, et~al.
\newblock Cytosolic {PLA}2 genes possibly contribute to the etiology of
  schizophrenia.
\newblock American Journal of Medical Genetics Part B: Neuropsychiatric
  Genetics. 2005;137(1):56--58.

\bibitem{nadalin2017ass}
Nadalin S, Bureti{\'c}-Tomljanovi{\'c} A.
\newblock An association between {PLA2G6} and {PLA2G4C} gene polymorphisms and
  schizophrenia risk and illness severity in a {C}roatian population.
\newblock Prostaglandins, Leukotrienes and Essential Fatty Acids.
  2017;121:57--59.

\bibitem{leHellard2010pol}
Le~Hellard S, M{\"u}hleisen T, Djurovic S, Fern{\o} J, Ouriaghi Z, Mattheisen
  M, et~al.
\newblock Polymorphisms in {SREBF}1 and {SREBF}2, two antipsychotic-activated
  transcription factors controlling cellular lipogenesis, are associated with
  schizophrenia in {G}erman and {S}candinavian samples.
\newblock Molecular psychiatry. 2010;15(5):463--472.

\bibitem{barbosa2007ass}
Barbosa NR, Junqueira RM, Vallada HP, Gattaz WF.
\newblock Association between {B}an {I} genotype and increased phospholipase
  {A}2 activity in schizophrenia.
\newblock European Archives of Psychiatry and Clinical Neuroscience.
  2007;257:340--343.

\bibitem{nadalin2008ban}
Nadalin S, Rube{\v{s}}a G, Giacometti J, Vulin M, Tomljanovi{\'c} D,
  Vranekovi{\'c} J, et~al.
\newblock Ban{I} polymorphism of cytosolic phospholipase {A}2 gene is
  associated with age at onset in male patients with schizophrenia and
  schizoaffective disorder.
\newblock Prostaglandins, leukotrienes and essential fatty acids.
  2008;78(6):351--360.

\bibitem{allison2009obe}
Allison DB, Newcomer JW, Dunn AL, Blumenthal JA, Fabricatore AN, Daumit GL,
  et~al.
\newblock Obesity among those with mental disorders: a National Institute of
  Mental Health meeting report.
\newblock American journal of preventive medicine. 2009;36(4):341--350.

\bibitem{garrido2022pre}
Garrido-Torres N, Ruiz-Veguilla M, Alameda L, Canal-Rivero M, Ruiz MJ,
  G{\'o}mez-Revuelta M, et~al.
\newblock Prevalence of metabolic syndrome and related factors in a large
  sample of antipsychotic na{\"\i}ve patients with first-episode psychosis:
  Baseline results from the {PAFIP} cohort.
\newblock Schizophrenia Research. 2022;246:277--285.

\bibitem{strassnig2005diereview}
Strassnig M, Brar J, Ganguli R.
\newblock Dietary intake of patients with schizophrenia.
\newblock Psychiatry (Edgmont (Pa: Township)). 2005;2(2):31--35.

\bibitem{strassnig2005die}
Strassnig M, Brar JS, Ganguli R.
\newblock Dietary fatty acid and antioxidant intake in community-dwelling
  patients suffering from schizophrenia.
\newblock Schizophrenia research. 2005;76(2-3):343--351.

\bibitem{dipasquale2013die}
Dipasquale S, Pariante CM, Dazzan P, Aguglia E, McGuire P, Mondelli V.
\newblock The dietary pattern of patients with schizophrenia: a systematic
  review.
\newblock Journal of psychiatric research. 2013;47(2):197--207.

\bibitem{pawelczyk2016ass}
Pawe{\l}czyk T, Trafalska E, Kotlicka-Antczak M, Pawe{\l}czyk A.
\newblock The association between polyunsaturated fatty acid consumption and
  the transition to psychosis in ultra-high risk individuals.
\newblock Prostaglandins, Leukotrienes and Essential Fatty Acids.
  2016;108:30--37.

\bibitem{pawelczyk2017dif}
Pawe{\l}czyk T, Trafalska E, Pawe{\l}czyk A, Kotlicka-Antczak M.
\newblock Differences in omega-3 and omega-6 polyunsaturated fatty acid
  consumption in people at ultra-high risk of psychosis, first-episode
  schizophrenia, and in healthy controls.
\newblock Early intervention in psychiatry. 2017;11(6):498--508.

\bibitem{kim2017low}
Kim EJ, Lim SY, Lee HJ, Lee JY, Choi S, Kim SY, et~al.
\newblock Low dietary intake of n-3 fatty acids, niacin, folate, and vitamin
  {C} in {K}orean patients with schizophrenia and the development of dietary
  guidelines for schizophrenia.
\newblock Nutrition Research. 2017;45:10--18.

\bibitem{mcnamara2009mod}
McNamara R.
\newblock Modulation of polyunsaturated fatty acid biosynthesis by
  antipsychotic medications: Implications for the pathophysiology and treatment
  of schizophrenia.
\newblock Clinical Lipidology. 2009;4(6):809--820.

\bibitem{bozzatello2019pol}
Bozzatello P, Rocca P, Mantelli E, Bellino S.
\newblock Polyunsaturated fatty acids: what is their role in treatment of
  psychiatric disorders?
\newblock International Journal of Molecular Sciences. 2019;20(21):5257.

\bibitem{hsu2020ben}
Hsu MC, Huang YS, Ouyang WC.
\newblock Beneficial effects of omega-3 fatty acid supplementation in
  schizophrenia: Possible mechanisms.
\newblock Lipids in Health and Disease. 2020;19:1--17.

\bibitem{kim2013fat}
Kim J, Li Y, Watkins BA.
\newblock Fat to treat fat: emerging relationship between dietary {PUFA},
  endocannabinoids, and obesity.
\newblock Prostaglandins \& other lipid mediators. 2013;104:32--41.

\bibitem{hedelin2010die}
Hedelin M, L{\"o}f M, Olsson M, Lewander T, Nilsson B, Hultman CM, et~al.
\newblock Dietary intake of fish, omega-3, omega-6 polyunsaturated fatty acids
  and vitamin {D} and the prevalence of psychotic-like symptoms in a cohort of
  33,000 women from the general population.
\newblock BMC psychiatry. 2010;10(1):1--13.

\bibitem{gao2023ass}
Gao Y, Hu X, Wang D, Jiang J, Li M, Qing Y, et~al.
\newblock Association between arachidonic acid and the risk of schizophrenia: a
  cross-national study and Mendelian randomization analysis.
\newblock Nutrients. 2023;15(5):1195.

\bibitem{fogerty1991com}
Fogerty A, Whitfield F, Svoronos D, Ford G.
\newblock The composition of the fatty acids and aldehydes of the ethanolamine
  and choline phospholipids of various meats.
\newblock International journal of food science \& technology.
  1991;26(4):363--371.

\bibitem{trepanier2016pos}
Tr{\'e}panier M, Hopperton K, Mizrahi R, Mechawar N, Bazinet R.
\newblock Postmortem evidence of cerebral inflammation in schizophrenia: a
  systematic review.
\newblock Molecular psychiatry. 2016;21(8):1009--1026.

\bibitem{miller2011met}
Miller BJ, Buckley P, Seabolt W, Mellor A, Kirkpatrick B.
\newblock Meta-analysis of cytokine alterations in schizophrenia: clinical
  status and antipsychotic effects.
\newblock Biological psychiatry. 2011;70(7):663--671.

\bibitem{khoury2018inf}
Khoury R, Nasrallah HA.
\newblock Inflammatory biomarkers in individuals at clinical high risk for
  psychosis ({CHR-P}): State or trait?
\newblock Schizophrenia research. 2018;199:31--38.

\bibitem{cullen2019ass}
Cullen AE, Holmes S, Pollak TA, Blackman G, Joyce DW, Kempton MJ, et~al.
\newblock Associations between non-neurological autoimmune disorders and
  psychosis: a meta-analysis.
\newblock Biological Psychiatry. 2019;85(1):35--48.

\bibitem{wang2018aut}
Wang LY, Chen SF, Chiang JH, Hsu CY, Shen YC.
\newblock Autoimmune diseases are associated with an increased risk of
  schizophrenia: a nationwide population-based cohort study.
\newblock Schizophrenia Research. 2018;202:297--302.

\bibitem{chen2009prev}
Chen YH, Lee HC, Lin HC.
\newblock Prevalence and risk of atopic disorders among schizophrenia patients:
  a nationwide population based study.
\newblock Schizophrenia research. 2009;108(1-3):191--196.

\bibitem{pedersen2012sch}
Pedersen MS, Benros ME, Agerbo E, B{\o}rglum AD, Mortensen PB.
\newblock Schizophrenia in patients with atopic disorders with particular
  emphasis on asthma: a {D}anish population-based study.
\newblock Schizophrenia Research. 2012;138(1):58--62.

\bibitem{tiosano2017sch}
Tiosano S, Farhi A, Watad A, Grysman N, Stryjer R, Amital H, et~al.
\newblock Schizophrenia among patients with systemic lupus erythematosus:
  population-based cross-sectional study.
\newblock Epidemiology and psychiatric sciences. 2017;26(4):424--429.

\bibitem{oken1999iss}
Oken RJ, Schulzer M.
\newblock At issue: schizophrenia and rheumatoid arthritis: the negative
  association revisited.
\newblock Schizophrenia Bulletin. 1999;25(4):625--638.

\bibitem{matthysse1975bio}
Matthysse S, Lipinski J.
\newblock Biochemical aspects of schizophrenia.
\newblock Annual review of medicine. 1975;26(1):551--565.

\bibitem{heleniak1999his}
Heleniak E, O'Desky I.
\newblock Histamine and prostaglandins in schizophrenia: revisited.
\newblock Medical hypotheses. 1999;52(1):37--42.

\bibitem{mccoy2002rol}
McCoy JM, Wicks JR, Audoly LP, et~al.
\newblock The role of prostaglandin {E2} receptors in the pathogenesis of
  rheumatoid arthritis.
\newblock The Journal of clinical investigation. 2002;110(5):651--658.

\bibitem{pettipher2008rol}
Pettipher R.
\newblock The roles of the prostaglandin {D2} receptors {DP1} and {CRTH2} in
  promoting allergic responses.
\newblock British journal of pharmacology. 2008;153(S1):S191--S199.

\bibitem{hallstrand2010up}
Hallstrand TS, Henderson~Jr WR.
\newblock An update on the role of leukotrienes in asthma.
\newblock Current opinion in allergy and clinical immunology. 2010;10(1):60.

\bibitem{ruzicka1986ski}
Ruzicka T, Simmet T, Peskar BA, Ring J.
\newblock Skin levels of arachidonic acid-derived inflammatory mediators and
  histamine in atopic dermatitis and psoriasis.
\newblock Journal of investigative dermatology. 1986;86(2):105--108.

\bibitem{horrobin1977sch}
Horrobin D.
\newblock Schizophrenia as a prostaglandin deficiency disease.
\newblock The Lancet. 1977;309(8018):936--937.

\bibitem{engels2014cli}
Engels G, Francke AL, van Meijel B, Douma JG, de~Kam H, Wesselink W, et~al.
\newblock Clinical pain in schizophrenia: a systematic review.
\newblock The Journal of Pain. 2014;15(5):457--467.

\bibitem{leweke1999ele}
Leweke FM, Giuffrida A, Wurster U, Emrich HM, Piomelli D.
\newblock Elevated endogenous cannabinoids in schizophrenia.
\newblock Neuroreport. 1999;10(8):1665--1669.

\bibitem{deMarchi2003end}
De~Marchi N, De~Petrocellis L, Orlando P, Daniele F, Fezza F, Di~Marzo V.
\newblock Endocannabinoid signalling in the blood of patients with
  schizophrenia.
\newblock Lipids in health and disease. 2003;2:1--9.

\bibitem{giuffrida2004cer}
Giuffrida A, Leweke FM, Gerth CW, Schreiber D, Koethe D, Faulhaber J, et~al.
\newblock Cerebrospinal anandamide levels are elevated in acute schizophrenia
  and are inversely correlated with psychotic symptoms.
\newblock Neuropsychopharmacology. 2004;29(11):2108--2114.

\bibitem{leweke2007ana}
Leweke FM, Giuffrida A, Koethe D, Schreiber D, Nolden BM, Kranaster L, et~al.
\newblock Anandamide levels in cerebrospinal fluid of first-episode
  schizophrenic patients: impact of cannabis use.
\newblock Schizophrenia research. 2007;94(1-3):29--36.

\bibitem{potvin2008end}
Potvin S, Kouassi {\'E}, Lipp O, Bouchard RH, Roy MA, Demers MF, et~al.
\newblock Endogenous cannabinoids in patients with schizophrenia and substance
  use disorder during quetiapine therapy.
\newblock Journal of psychopharmacology. 2008;22(3):262--269.

\bibitem{koethe2009ana}
Koethe D, Giuffrida A, Schreiber D, Hellmich M, Schultze-Lutter F, Ruhrmann S,
  et~al.
\newblock Anandamide elevation in cerebrospinal fluid in initial prodromal
  states of psychosis.
\newblock The British Journal of Psychiatry. 2009;194(4):371--372.

\bibitem{muguruza2013qua}
Muguruza C, Lehtonen M, Aaltonen N, Morentin B, Meana JJ, Callado LF.
\newblock Quantification of endocannabinoids in postmortem brain of
  schizophrenic subjects.
\newblock Schizophrenia Research. 2013;148(1-3):145--150.

\bibitem{koethe2019fam}
Koethe D, Pahlisch F, Hellmich M, Rohleder C, Mueller JK, Meyer-Lindenberg A,
  et~al.
\newblock Familial abnormalities of endocannabinoid signaling in schizophrenia.
\newblock The World Journal of Biological Psychiatry. 2019;20(2):117--125.

\bibitem{potvin2020per}
Potvin S, Mahrouche L, Assaf R, Chicoine M, Gigu{\`e}re C{\'E}, Furtos A,
  et~al.
\newblock Peripheral endogenous cannabinoid levels are increased in
  schizophrenia patients evaluated in a psychiatric emergency setting.
\newblock Frontiers in Psychiatry. 2020;11:628.

\bibitem{ibarra2022can}
Ibarra-Lecue I, Unzueta-Larrinaga P, Barrena-Barbadillo R, Villate A, Horrillo
  I, Mendivil B, et~al.
\newblock Cannabis use selectively modulates circulating biomarkers in the
  blood of schizophrenia patients.
\newblock Addiction biology. 2022;27(6):e13233.

\bibitem{parksepp2022exp}
Parksepp M, Haring L, Kilk K, Koch K, Uppin K, Kangro R, et~al.
\newblock The expanded endocannabinoid system contributes to metabolic and body
  mass shifts in first-episode schizophrenia: a 5-year follow-up study.
\newblock Biomedicines. 2022;10(2):243.

\bibitem{romer2023bio}
R{\o}mer TB, Jeppesen R, Christensen RHB, Benros ME.
\newblock Biomarkers in the cerebrospinal fluid of patients with psychotic
  disorders compared to healthy controls: a systematic review and
  meta-analysis.
\newblock Molecular Psychiatry. 2023;p. 1--14.

\bibitem{minichino2019mea}
Minichino A, Senior M, Brondino N, Zhang SH, Godlewska BR, Burnet PW, et~al.
\newblock Measuring disturbance of the endocannabinoid system in psychosis: a
  systematic review and meta-analysis.
\newblock JAMA psychiatry. 2019;76(9):914--923.

\bibitem{appiah2020chi}
Appiah-Kusi E, Wilson R, Colizzi M, Foglia E, Klamerus E, Caldwell A, et~al.
\newblock Childhood trauma and being at-risk for psychosis are associated with
  higher peripheral endocannabinoids.
\newblock Psychological medicine. 2020;50(11):1862--1871.

\bibitem{joaquim2022pla}
Joaquim HP, Costa AC, Pereira CA, Talib LL, Bilt MM, Loch AA, et~al.
\newblock Plasmatic endocannabinoids are decreased in subjects with ultra-high
  risk of psychosis.
\newblock European Journal of Neuroscience. 2022;55(4):1079--1087.

\bibitem{muguruza2019end}
Muguruza C, Morentin B, Meana JJ, Alexander SP, Callado LF.
\newblock Endocannabinoid system imbalance in the postmortem prefrontal cortex
  of subjects with schizophrenia.
\newblock Journal of Psychopharmacology. 2019;33(9):1132--1140.

\bibitem{bioque2013per}
Bioque M, Garc{\'\i}a-Bueno B, MacDowell KS, Meseguer A, Saiz PA, Parellada M,
  et~al.
\newblock Peripheral endocannabinoid system dysregulation in first-episode
  psychosis.
\newblock Neuropsychopharmacology. 2013;38(13):2568--2577.

\bibitem{morgan2013cer}
Morgan CJ, Page E, Schaefer C, Chatten K, Manocha A, Gulati S, et~al.
\newblock Cerebrospinal fluid anandamide levels, cannabis use and
  psychotic-like symptoms.
\newblock The British Journal of Psychiatry. 2013;202(5):381--382.

\bibitem{si2018ass}
Si P, Liu S, Tong D, Cheng M, Wang L, Cheng X.
\newblock Association of polymorphisms of {NAPE-PLD} and {FAAH} genes with
  schizophrenia in {C}hinese {H}an population.
\newblock Zhonghua yi xue yi chuan xue za zhi= Zhonghua yixue yichuanxue zazhi=
  Chinese journal of medical genetics. 2018;35(2):215--218.

\bibitem{wood2019tar}
Wood PL.
\newblock Targeted lipidomics and metabolomics evaluations of cortical neuronal
  stress in schizophrenia.
\newblock Schizophrenia Research. 2019;212:107--112.

\bibitem{dean2001stu}
Bradbury R, Copolov D.
\newblock Studies on {[3H] CP-55940} binding in the human central nervous
  system: regional specific changes in density of cannabinoid-1 receptors
  associated with schizophrenia and cannabis use.
\newblock Neuroscience. 2001;103(1):9--15.

\bibitem{zavitsanou2004sel}
Zavitsanou K, Garrick T, Huang XF.
\newblock Selective antagonist {[3H] SR141716A} binding to cannabinoid {CB1}
  receptors is increased in the anterior cingulate cortex in schizophrenia.
\newblock Progress in Neuro-Psychopharmacology and Biological Psychiatry.
  2004;28(2):355--360.

\bibitem{newell2006inc}
Newell KA, Deng C, Huang XF.
\newblock Increased cannabinoid receptor density in the posterior cingulate
  cortex in schizophrenia.
\newblock Experimental brain research. 2006;172:556--560.

\bibitem{wong2010qua}
Wong DF, Kuwabara H, Horti AG, Raymont V, Brasic J, Guevara M, et~al.
\newblock Quantification of cerebral cannabinoid receptors subtype 1 ({CB1}) in
  healthy subjects and schizophrenia by the novel {PET} radioligand {[11C]
  OMAR}.
\newblock Neuroimage. 2010;52(4):1505--1513.

\bibitem{jenko2012bin}
Jenko KJ, Hirvonen J, Henter ID, Anderson KB, Zoghbi SS, Hyde TM, et~al.
\newblock Binding of a tritiated inverse agonist to cannabinoid {CB1} receptors
  is increased in patients with schizophrenia.
\newblock Schizophrenia research. 2012;141(2-3):185--188.

\bibitem{ceccarini2013inc}
Ceccarini J, De~Hert M, Van~Winkel R, Peuskens J, Bormans G, Kranaster L,
  et~al.
\newblock Increased ventral striatal {CB1} receptor binding is related to
  negative symptoms in drug-free patients with schizophrenia.
\newblock Neuroimage. 2013;79:304--312.

\bibitem{volk2014rec}
Volk DW, Eggan SM, Horti AG, Wong DF, Lewis DA.
\newblock Reciprocal alterations in cortical cannabinoid receptor 1 binding
  relative to protein immunoreactivity and transcript levels in schizophrenia.
\newblock Schizophrenia research. 2014;159(1):124--129.

\bibitem{chase2016cha}
Chase KA, Feiner B, Rosen C, Gavin DP, Sharma RP.
\newblock Characterization of peripheral cannabinoid receptor expression and
  clinical correlates in schizophrenia.
\newblock Psychiatry Research. 2016;245:346--353.

\bibitem{tao2020can}
Tao R, Li C, Jaffe AE, Shin JH, Deep-Soboslay A, Yamin R, et~al.
\newblock Cannabinoid receptor {CNR1} expression and {DNA} methylation in human
  prefrontal cortex, hippocampus and caudate in brain development and
  schizophrenia.
\newblock Translational psychiatry. 2020;10(1):158.

\bibitem{chou2023ter}
Chou S, Fish KN, Lewis DA, Sweet RA.
\newblock Terminal type-specific cannabinoid {CB1} receptor alterations in
  patients with schizophrenia: a pilot study.
\newblock bioRxiv. 2023;p. 2023--04.

\bibitem{dalton2011par}
Dalton VS, Long LE, Weickert CS, Zavitsanou K.
\newblock Paranoid schizophrenia is characterized by increased {CB1} receptor
  binding in the dorsolateral prefrontal cortex.
\newblock Neuropsychopharmacology. 2011;36(8):1620--1630.

\bibitem{eggan2008red}
Eggan SM, Hashimoto T, Lewis DA.
\newblock Reduced cortical cannabinoid 1 receptor messenger {RNA} and protein
  expression in schizophrenia.
\newblock Archives of general psychiatry. 2008;65(7):772--784.

\bibitem{uriguen2009imm}
Urig{\"u}en L, Garc{\'\i}a-Fuster MJ, Callado LF, Morentin B, La~Harpe R,
  Casad{\'o} V, et~al.
\newblock Immunodensity and m{RNA} expression of {A2A} adenosine, {D2}
  dopamine, and {CB1} cannabinoid receptors in postmortem frontal cortex of
  subjects with schizophrenia: effect of antipsychotic treatment.
\newblock Psychopharmacology. 2009;206:313--324.

\bibitem{eggan2010can}
Eggan SM, Stoyak SR, Verrico CD, Lewis DA.
\newblock Cannabinoid {CB1} receptor immunoreactivity in the prefrontal cortex:
  comparison of schizophrenia and major depressive disorder.
\newblock Neuropsychopharmacology. 2010;35(10):2060--2071.

\bibitem{ranganathan2016red}
Ranganathan M, Cortes-Briones J, Radhakrishnan R, Thurnauer H, Planeta B,
  Skosnik P, et~al.
\newblock Reduced brain cannabinoid receptor availability in schizophrenia.
\newblock Biological psychiatry. 2016;79(12):997--1005.

\bibitem{borgan2019viv}
Borgan F, Laurikainen H, Veronese M, Marques TR, Haaparanta-Solin M, Solin O,
  et~al.
\newblock In vivo availability of cannabinoid 1 receptor levels in patients
  with first-episode psychosis.
\newblock JAMA psychiatry. 2019;76(10):1074--1084.

\bibitem{deng2007no}
Deng C, Han M, Huang XF.
\newblock No changes in densities of cannabinoid receptors in the superior
  temporal gyrus in schizophrenia.
\newblock Neuroscience bulletin. 2007;23(6):341.

\bibitem{koethe2007exp}
Koethe D, Llenos I, Dulay J, Hoyer C, Torrey E, Leweke F, et~al.
\newblock Expression of {CB1} cannabinoid receptor in the anterior cingulate
  cortex in schizophrenia, bipolar disorder, and major depression.
\newblock Journal of neural transmission. 2007;114:1055--1063.

\bibitem{wallace1993tra}
Wallace MA, Claro E.
\newblock Transmembrane signaling through phospholipase {C} in human cortical
  membranes.
\newblock Neurochemical research. 1993;18:139--145.

\bibitem{dAddario2017pre}
D'Addario C, Micale V, Di~Bartolomeo M, Stark T, Pucci M, Sulcova A, et~al.
\newblock A preliminary study of endocannabinoid system regulation in
  psychosis: Distinct alterations of {CNR1} promoter {DNA} methylation in
  patients with schizophrenia.
\newblock Schizophrenia research. 2017;188:132--140.

\bibitem{navarro2022mol}
Navarro D, Gasparyan A, Navarrete F, Torregrosa AB, Rubio G, Mar{\'\i}n-Mayor
  M, et~al.
\newblock Molecular alterations of the endocannabinoid system in psychiatric
  disorders.
\newblock International journal of molecular sciences. 2022;23(9):4764.

\bibitem{covey2017end}
Covey DP, Mateo Y, Sulzer D, Cheer JF, Lovinger DM.
\newblock Endocannabinoid modulation of dopamine neurotransmission.
\newblock Neuropharmacology. 2017;124:52--61.

\bibitem{carvalho2012can}
Carvalho AF, Van~Bockstaele EJ.
\newblock Cannabinoid modulation of noradrenergic circuits: implications for
  psychiatric disorders.
\newblock Progress in Neuro-Psychopharmacology and Biological Psychiatry.
  2012;38(1):59--67.

\bibitem{haj2011mod}
Haj-Dahmane S, Shen RY.
\newblock Modulation of the serotonin system by endocannabinoid signaling.
\newblock Neuropharmacology. 2011;61(3):414--420.

\bibitem{maia2017int}
Maia TV, Frank MJ.
\newblock An integrative perspective on the role of dopamine in schizophrenia.
\newblock Biological psychiatry. 2017;81(1):52--66.

\bibitem{juckel2015ser}
Juckel G.
\newblock Serotonin: from sensory processing to schizophrenia using an
  electrophysiological method.
\newblock Behavioural brain research. 2015;277:121--124.

\bibitem{harkany2003com}
Harkany T, H{\"a}rtig W, Berghuis P, Dobszay MB, Zilberter Y, Edwards RH,
  et~al.
\newblock Complementary distribution of type 1 cannabinoid receptors and
  vesicular glutamate transporter 3 in basal forebrain suggests input-specific
  retrograde signalling by cholinergic neurons.
\newblock European Journal of Neuroscience. 2003;18(7):1979--1992.

\bibitem{tzavara2003bip}
Tzavara ET, Wade M, Nomikos GG.
\newblock Biphasic effects of cannabinoids on acetylcholine release in the
  hippocampus: site and mechanism of action.
\newblock Journal of Neuroscience. 2003;23(28):9374--9384.

\bibitem{saunders2015cor}
Saunders A, Granger AJ, Sabatini BL.
\newblock Corelease of acetylcholine and {GABA} from cholinergic forebrain
  neurons.
\newblock Elife. 2015;4:e06412.

\bibitem{gibbons2016cho}
Gibbons A, Dean B.
\newblock The cholinergic system: an emerging drug target for schizophrenia.
\newblock Current pharmaceutical design. 2016;22(14):2124--2133.

\bibitem{robinson2010win}
Robinson L, Goonawardena AV, Pertwee R, Hampson RE, Platt B, Riedel G.
\newblock {WIN55}, 212-2 induced deficits in spatial learning are mediated by
  cholinergic hypofunction.
\newblock Behavioural Brain Research. 2010;208(2):584--592.

\bibitem{navakkode2014pha}
Navakkode S, Korte M.
\newblock Pharmacological activation of {CB1} receptor modulates long term
  potentiation by interfering with protein synthesis.
\newblock Neuropharmacology. 2014;79:525--533.

\bibitem{gonzalez2016pro}
Gonzalez-Blanco L, Greenhalgh AMD, Garcia-Rizo C, Fernandez-Egea E, Miller BJ,
  Kirkpatrick B.
\newblock Prolactin concentrations in antipsychotic-naive patients with
  schizophrenia and related disorders: a meta-analysis.
\newblock Schizophrenia research. 2016;174(1-3):156--160.

\bibitem{labad2015str}
Labad J, Stojanovic-P{\'e}rez A, Montalvo I, Sol{\'e} M, Cabezas {\'A}, Ortega
  L, et~al.
\newblock Stress biomarkers as predictors of transition to psychosis in at-risk
  mental states: roles for cortisol, prolactin and albumin.
\newblock Journal of psychiatric research. 2015;60:163--169.

\bibitem{grandison1984sti}
GRANDISON L.
\newblock Stimulation of anterior pituitary prolactin release by melittin, an
  activator of phospholipase {A2}.
\newblock Endocrinology. 1984;114(1):1--7.

\bibitem{ross1988dyn}
ROSS PC, JUDD AM, MACLEOD RM.
\newblock The dynamics of arachidonic acid liberation and prolactin release: a
  comparison of thyrotropin-releasing hormone, angiotensin {II}, and
  neurotensin stimulation in perifused rat anterior pituitary cells.
\newblock Endocrinology. 1988;123(5):2445--2453.

\bibitem{pruessner2017neu}
Pruessner M, Cullen AE, Aas M, Walker EF.
\newblock The neural diathesis-stress model of schizophrenia revisited: an
  update on recent findings considering illness stage and neurobiological and
  methodological complexities.
\newblock Neuroscience \& Biobehavioral Reviews. 2017;73:191--218.

\bibitem{hashimoto2008alt}
Unger T, Maldonado-Aviles J, Morris H, Volk D, Mirnics K, Lewis D.
\newblock Alterations in {GABA}-related transcriptome in the dorsolateral
  prefrontal cortex of subjects with schizophrenia.
\newblock Molecular psychiatry. 2008;13(2):147--161.

\bibitem{hammond2014evi}
Hammond J, Shan D, Meador-Woodruff J, McCullumsmith R.
\newblock Evidence of glutamatergic dysfunction in the pathophysiology of
  schizophrenia.
\newblock Synaptic stress and pathogenesis of neuropsychiatric disorders.
  2014;p. 265--294.

\bibitem{friston2016dys}
Friston K, Brown HR, Siemerkus J, Stephan KE.
\newblock The dysconnection hypothesis (2016).
\newblock Schizophrenia research. 2016;176(2-3):83--94.

\bibitem{peralta2010dsm}
Peralta V, Campos MS, de~Jalon EG, Cuesta MJ.
\newblock {DSM-IV} catatonia signs and criteria in first-episode, drug-naive,
  psychotic patients: psychometric validity and response to antipsychotic
  medication.
\newblock Schizophrenia research. 2010;118(1-3):168--175.

\bibitem{cuesta2018mot}
Cuesta MJ, de~Jal{\'o}n EG, Campos MS, Moreno-Izco L, Lorente-Ome{\~n}aca R,
  S{\'a}nchez-Torres AM, et~al.
\newblock Motor abnormalities in first-episode psychosis patients and long-term
  psychosocial functioning.
\newblock Schizophrenia research. 2018;200:97--103.

\bibitem{pinkham2015amy}
Pinkham AE, Liu P, Lu H, Kriegsman M, Simpson C, Tamminga C.
\newblock Amygdala hyperactivity at rest in paranoid individuals with
  schizophrenia.
\newblock American Journal of Psychiatry. 2015;172(8):784--792.

\bibitem{taylor2005neu}
Taylor SF, Luan~Phan K, Britton JC, Liberzon I.
\newblock Neural response to emotional salience in schizophrenia.
\newblock Neuropsychopharmacology. 2005;30(5):984--995.

\bibitem{potvin2016emo}
Potvin S, Tik{\`a}sz A, Mendrek A.
\newblock Emotionally neutral stimuli are not neutral in schizophrenia: a mini
  review of functional neuroimaging studies.
\newblock Frontiers in psychiatry. 2016;7:115.

\bibitem{matthews2014vis}
Matthews NL, Collins KP, Thakkar KN, Park S.
\newblock Visuospatial imagery and working memory in schizophrenia.
\newblock Cognitive neuropsychiatry. 2014;19(1):17--35.

\bibitem{bodatsch2015for}
Bodatsch M, Brockhaus-Dumke A, Klosterk{\"o}tter J, Ruhrmann S.
\newblock Forecasting psychosis by event-related potentials--systematic review
  and specific meta-analysis.
\newblock Biological Psychiatry. 2015;77(11):951--958.

\bibitem{berkovitch2018imp}
Berkovitch L, Del~Cul A, Maheu M, Dehaene S.
\newblock Impaired conscious access and abnormal attentional amplification in
  schizophrenia.
\newblock NeuroImage: Clinical. 2018;18:835--848.

\bibitem{wynn2010imp}
Wynn JK, Horan WP, Kring AM, Simons RF, Green MF.
\newblock Impaired anticipatory event-related potentials in schizophrenia.
\newblock International Journal of Psychophysiology. 2010;77(2):141--149.

\bibitem{skosnik2016its}
Skosnik PD, Cortes-Briones JA, Haj{\'o}s M.
\newblock It's all in the rhythm: the role of cannabinoids in neural
  oscillations and psychosis.
\newblock Biological psychiatry. 2016;79(7):568--577.

\bibitem{uhlhaas2015osc}
Uhlhaas PJ, Singer W.
\newblock Oscillations and neuronal dynamics in schizophrenia: the search for
  basic symptoms and translational opportunities.
\newblock Biological psychiatry. 2015;77(12):1001--1009.

\bibitem{lopes2015ang}
Lopes R, Soares R, Coelho R, Figueiredo-Braga M.
\newblock Angiogenesis in the pathophysiology of schizophrenia--a comprehensive
  review and a conceptual hypothesis.
\newblock Life sciences. 2015;128:79--93.

\bibitem{misir2023syn}
M{\i}s{\i}r E, Akay GG.
\newblock Synaptic dysfunction in schizophrenia.
\newblock Synapse;p. e22276.

\bibitem{mouchlis2016dev}
Mouchlis VD, Limnios D, Kokotou MG, Barbayianni E, Kokotos G, McCammon JA,
  et~al.
\newblock Development of potent and selective inhibitors for group {VIA}
  calcium-independent phospholipase {A}2 guided by molecular dynamics and
  structure--activity relationships.
\newblock Journal of medicinal chemistry. 2016;59(9):4403--4414.

\bibitem{dedaki2019bet}
Dedaki C, Kokotou MG, Mouchlis VD, Limnios D, Lei X, Mu CT, et~al.
\newblock $\beta$-Lactones: A novel class of {C}a2+-independent phospholipase
  {A}2 (group {VIA} i{PLA}2) inhibitors with the ability to inhibit
  $\beta$-cell apoptosis.
\newblock Journal of medicinal chemistry. 2019;62(6):2916--2927.

\bibitem{smyrniotou2017oxo}
Smyrniotou A, Kokotou MG, Mouchlis VD, Barbayianni E, Kokotos G, Dennis EA,
  et~al.
\newblock 2-Oxoamides based on dipeptides as selective calcium-independent
  phospholipase {A}2 inhibitors.
\newblock Bioorganic \& medicinal chemistry. 2017;25(3):926--940.

\bibitem{dubin1982pha}
Dubin NH, Blake DA, DiBlasi MC, Parmley TH, King TM.
\newblock Pharmacokinetic studies on quinacrine following intrauterine
  administration to cynomolgus monkeys.
\newblock Fertility and Sterility. 1982;38(6):735--740.

\bibitem{nevin2016psy}
Nevin RL, Croft AM.
\newblock Psychiatric effects of malaria and anti-malarial drugs: historical
  and modern perspectives.
\newblock Malaria Journal. 2016;15(1):1--13.

\bibitem{oDonnell1991mus}
O'Donnell KA, Howlett AC.
\newblock Muscarinic receptor binding is inhibited by quinacrine.
\newblock Neuroscience letters. 1991;127(1):46--48.

\bibitem{tamamizu1995mut}
Tamamizu S, Todd AP, McNamee MG.
\newblock Mutations in the {M}1 region of the nicotinic acetylcholine receptor
  alter the sensitivity to inhibition by quinacrine.
\newblock Cellular and molecular neurobiology. 1995;15:427--438.

\bibitem{lidz1946tox}
Lidz T, Kahn RL.
\newblock Toxicity of quinacrine (atabrine) for the central nervous system:
  {III. A}n experimental study on human subjects.
\newblock Archives of Neurology \& Psychiatry. 1946;56(3):284--299.

\bibitem{newell1946tox}
NEWELL HW, LIDZ T.
\newblock The toxicity of atabrine to the central nervous system.
\newblock American Journal of Psychiatry. 1946;102(6):805--818.

\bibitem{kitagawa2021ant}
Kitagawa T, Matsumoto A, Terashima I, Uesono Y.
\newblock Antimalarial quinacrine and chloroquine lose their activity by
  decreasing cationic amphiphilic structure with a slight decrease in p{H}.
\newblock Journal of Medicinal Chemistry. 2021;64(7):3885--3896.

\bibitem{vadas1986pot}
Vadas P, Stefanski E, Pruzanski W.
\newblock Potential therapeutic efficacy of inhibitors of human phospholipase
  {A}2 in septic shock.
\newblock Agents and Actions. 1986;19(3-4):194--202.

\bibitem{christensen2007eff}
Christensen R, Kristensen PK, Bartels EM, Bliddal H, Astrup A.
\newblock Efficacy and safety of the weight-loss drug rimonabant: a
  meta-analysis of randomised trials.
\newblock The Lancet. 2007;370(9600):1706--1713.

\bibitem{kelly2011eff}
Kelly DL, Gorelick DA, Conley RR, Boggs DL, Linthicum J, Liu F, et~al.
\newblock Effects of the cannabinoid-1 receptor antagonist rimonabant on
  psychiatric symptoms in overweight people with schizophrenia: a randomized,
  double-blind, pilot study.
\newblock Journal of clinical psychopharmacology. 2011;31(1):86.

\bibitem{meltzer2004pla}
Meltzer HY, Arvanitis L, Bauer D, Rein W, Group MTS.
\newblock Placebo-controlled evaluation of four novel compounds for the
  treatment of schizophrenia and schizoaffective disorder.
\newblock American Journal of Psychiatry. 2004;161(6):975--984.

\bibitem{sanofi2009SZ}
Sanofi. Efficacy and Safety of {AVE1625} as a Co-treatment With Antipsychotic
  Therapy in Schizophrenia; 2009.
\newblock Accessed: 2023-06-28.
\newblock http:// ClinicalTrials.gov/ show/ NCT00439634.

\bibitem{iseger2015sys}
Iseger TA, Bossong MG.
\newblock A systematic review of the antipsychotic properties of cannabidiol in
  humans.
\newblock Schizophrenia research. 2015;162(1-3):153--161.

\bibitem{xu2022eff}
Xu X, Shao G, Zhang X, Hu Y, Huang J, Su Y, et~al.
\newblock The efficacy of nutritional supplements for the adjunctive treatment
  of schizophrenia in adults: a systematic review and network meta-analysis.
\newblock Psychiatry Research. 2022;311:114500.

\bibitem{chen1996oxi}
Chen X, Gresham A, Morrison A, Pentland AP.
\newblock Oxidative stress mediates synthesis of cytosolic phospholipase {A2}
  after {UVB} injury.
\newblock Biochimica et Biophysica Acta (BBA)-Lipids and Lipid Metabolism.
  1996;1299(1):23--33.

\bibitem{quintana2019D2}
Quintana C, Beaulieu JM.
\newblock A fresh look at cortical dopamine {D2} receptor expressing neurons.
\newblock Pharmacol Res. 2019;139:440--445.

\bibitem{deLaMora2012dis}
De~La~Mora MP, Gallegos-Cari A, Crespo-Ramirez M, Marcellino D, Hansson A, Fuxe
  K.
\newblock Distribution of dopamine {D2}-like receptors in the rat amygdala and
  their role in the modulation of unconditioned fear and anxiety.
\newblock Neuroscience. 2012;201:252--266.

\bibitem{giuffrida1999dop}
Giuffrida A, Parsons L, Kerr T, De~Fonseca FR, Navarro M, Piomelli D.
\newblock Dopamine activation of endogenous cannabinoid signaling in dorsal
  striatum.
\newblock Nature neuroscience. 1999;2(4):358--363.

\bibitem{patel2003dif}
Patel S, Rademacher DJ, Hillard CJ.
\newblock Differential regulation of the endocannabinoids anandamide and
  2-arachidonylglycerol within the limbic forebrain by dopamine receptor
  activity.
\newblock Journal of Pharmacology and Experimental Therapeutics.
  2003;306(3):880--888.

\bibitem{centonze2004cri}
Centonze D, Battista N, Rossi S, Mercuri NB, Finazzi-Agro A, Bernardi G, et~al.
\newblock A critical interaction between dopamine {D}2 receptors and
  endocannabinoids mediates the effects of cocaine on striatal {GABA}ergic
  transmission.
\newblock Neuropsychopharmacology. 2004;29(8):1488--1497.

\bibitem{kerr2013ant}
Kerr DS, Talib LL, Yamamoto VJ, Ferreira AS, Zanetti MV, Serpa MH, et~al.
\newblock Antipsychotic drugs decrease i{PLA2} gene expression in
  schizophrenia.
\newblock Schizophrenia research. 2013;147(1):203--204.

\bibitem{yuhas2022clo}
Yuhas Y, Ashkenazi S, Berent E, Weizman A.
\newblock Clozapine Suppresses the Gene Expression and the Production of
  Cytokines and Up-Regulates Cyclooxygenase 2 m{RNA} in Human Astroglial Cells.
\newblock Brain Sciences. 2022;12(12):1703.

\bibitem{modi2013chr}
Modi HR, Taha AY, Kim HW, Chang L, Rapoport SI, Cheon Y.
\newblock Chronic clozapine reduces rat brain arachidonic acid metabolism by
  reducing plasma arachidonic acid availability.
\newblock Journal of neurochemistry. 2013;124(3):376--387.

\bibitem{felder1991tra}
Felder C, Williams H, Axelrod J.
\newblock A transduction pathway associated with receptors coupled to the
  inhibitory guanine nucleotide binding protein {G}i that amplifies
  {ATP}-mediated arachidonic acid release.
\newblock Proceedings of the National Academy of Sciences.
  1991;88(15):6477--6480.

\bibitem{piomelli1991dop}
Piomelli D, Pilon C, Giros B, Sokolofff P, Martres MP, Schwartz JC.
\newblock Dopamine activation of the arachidonic acid cascade as a basis for
  {D1/D2} receptor synergism.
\newblock Nature. 1991;353(6340):164--167.

\bibitem{felder1990ser}
Felder C, Kanterman R, Ma A, Axelrod J.
\newblock Serotonin stimulates phospholipase {A2} and the release of
  arachidonic acid in hippocampal neurons by a type 2 serotonin receptor that
  is independent of inositolphospholipid hydrolysis.
\newblock Proceedings of the National Academy of Sciences.
  1990;87(6):2187--2191.

\bibitem{dzitoyeva2013lox}
Dzitoyeva S, Chen H, Manev H, et~al.
\newblock 5-lipoxygenase-activating protein as a modulator of
  olanzapine-induced lipid accumulation in adipocyte.
\newblock Journal of lipids;2013.

\bibitem{raeder2006sre}
Raeder MB, Fern{\o} J, Vik-Mo AO, Steen VM.
\newblock {SREBP} activation by antipsychotic-and antidepressant-drugs in
  cultured human liver cells: relevance for metabolic side-effects?
\newblock Molecular and cellular biochemistry. 2006;289:167--173.

\bibitem{polymeropoulos2009com}
Polymeropoulos MH, Licamele L, Volpi S, Mack K, Mitkus SN, Carstea ED, et~al.
\newblock Common effect of antipsychotics on the biosynthesis and regulation of
  fatty acids and cholesterol supports a key role of lipid homeostasis in
  schizophrenia.
\newblock Schizophrenia research. 2009;108(1-3):134--142.

\bibitem{mcnamara2011dif}
McNamara RK, Jandacek R, Rider T, Tso P, Cole-Strauss A, Lipton JW.
\newblock Differential effects of antipsychotic medications on polyunsaturated
  fatty acid biosynthesis in rats: Relationship with liver delta6-desaturase
  expression.
\newblock Schizophrenia Research. 2011;129(1):57--65.

\bibitem{kim2012eff}
Kim HW, Cheon Y, Modi HR, Rapoport SI, Rao JS.
\newblock Effects of chronic clozapine administration on markers of arachidonic
  acid cascade and synaptic integrity in rat brain.
\newblock Psychopharmacology. 2012;222:663--674.

\bibitem{cheon2011chr}
Cheon Y, Park JY, Modi HR, Kim HW, Lee HJ, Chang L, et~al.
\newblock Chronic olanzapine treatment decreases arachidonic acid turnover and
  prostaglandin {E2} concentration in rat brain.
\newblock Journal of neurochemistry. 2011;119(2):364--376.

\bibitem{wu2016com}
Wu CS, Wang SC, Yeh IJ, Liu SK.
\newblock Comparative risk of seizure with use of first-and second-generation
  antipsychotics in patients with schizophrenia and mood disorders.
\newblock The Journal of clinical psychiatry. 2016;77(5):3433.

\bibitem{kearn2005con}
Kearn CS, Blake-Palmer K, Daniel E, Mackie K, Glass M.
\newblock Concurrent stimulation of cannabinoid {CB1} and dopamine {D2}
  receptors enhances heterodimer formation: a mechanism for receptor
  cross-talk?
\newblock Molecular pharmacology. 2005;67(5):1697--1704.

\bibitem{bagher2016ant}
Bagher AM, Laprairie RB, Kelly ME, Denovan-Wright EM.
\newblock Antagonism of dopamine receptor 2 long affects cannabinoid receptor 1
  signaling in a cell culture model of striatal medium spiny projection
  neurons.
\newblock Molecular Pharmacology. 2016;89(6):652--666.

\bibitem{guinart2020alt}
Guinart D, Moreno E, Galindo L, Cuenca-Royo A, Barrera-Conde M, P{\'e}rez EJ,
  et~al.
\newblock Altered signaling in {CB1R-5-HT2AR} heteromers in olfactory
  neuroepithelium cells of schizophrenia patients is modulated by cannabis use.
\newblock Schizophrenia Bulletin. 2020;46(6):1547--1557.

\bibitem{carhart2017ser}
Carhart-Harris R, Nutt D.
\newblock Serotonin and brain function: a tale of two receptors.
\newblock Journal of Psychopharmacology. 2017;31(9):1091--1120.

\bibitem{chew2008ant}
Chew ML, Mulsant BH, Pollock BG, Lehman ME, Greenspan A, Mahmoud RA, et~al.
\newblock Anticholinergic activity of 107 medications commonly used by older
  adults.
\newblock Journal of the American Geriatrics Society. 2008;56(7):1333--1341.

\bibitem{bordia2016str}
Bordia T, Zhang D, Perez XA, Quik M.
\newblock Striatal cholinergic interneurons and {D2} receptor-expressing
  {GABA}ergic medium spiny neurons regulate tardive dyskinesia.
\newblock Experimental neurology. 2016;286:32--39.

\bibitem{kharkwal2016PD}
Kharkwal G, Brami-Cherrier K, Lizardi-Ortiz JE, Nelson AB, Ramos M, Del~Barrio
  D, et~al.
\newblock Parkinsonism driven by antipsychotics originates from dopaminergic
  control of striatal cholinergic interneurons.
\newblock Neuron. 2016;91(1):67--78.

\bibitem{harrow2022twe}
Harrow M, Jobe TH, Tong L.
\newblock Twenty-year effects of antipsychotics in schizophrenia and affective
  psychotic disorders.
\newblock Psychological medicine. 2022;52(13):2681--2691.

\bibitem{moncrieff2023ant}
Moncrieff J, Crellin N, Stansfeld J, Cooper R, Marston L, Freemantle N, et~al.
\newblock Antipsychotic dose reduction and discontinuation versus maintenance
  treatment in people with schizophrenia and other recurrent psychotic
  disorders in England (the {RADAR} trial): an open, parallel-group, randomised
  controlled trial.
\newblock The Lancet Psychiatry. 2023;p. S2215--0366.

\bibitem{cai2022dim}
Cai H, Zeng C, Zhang X, Liu Y, Wu R, Guo W, et~al.
\newblock Diminished treatment response in relapsed versus first-episode
  schizophrenia as revealed by a panel of blood-based biomarkers: A combined
  cross-sectional and longitudinal study.
\newblock Psychiatry Research. 2022;316:114762.

\bibitem{francey2020psy}
Francey SM, O'Donoghue B, Nelson B, Graham J, Baldwin L, Yuen HP, et~al.
\newblock Psychosocial intervention with or without antipsychotic medication
  for first-episode psychosis: a randomized noninferiority clinical trial.
\newblock Schizophrenia Bulletin Open. 2020;1(1):sgaa015.

\bibitem{desfosses2010end}
Desfoss{\'e}s J, Stip E, Bentaleb LA, Potvin S.
\newblock Endocannabinoids and schizophrenia.
\newblock Pharmaceuticals. 2010;3(10):3101--3126.

\bibitem{cohen2010emo}
Cohen AS, Minor KS.
\newblock Emotional experience in patients with schizophrenia revisited:
  meta-analysis of laboratory studies.
\newblock Schizophrenia bulletin. 2010;36(1):143--150.

\bibitem{kring2010emo}
Kring AM, Caponigro JM.
\newblock Emotion in schizophrenia: where feeling meets thinking.
\newblock Current directions in psychological science. 2010;19(4):255--259.

\bibitem{yan2019ant}
Yan C, Lui SS, Zou Lq, Wang Cy, Zhou Fc, Cheung EF, et~al.
\newblock Anticipatory pleasure for future rewards is attenuated in patients
  with schizophrenia but not in individuals with schizotypal traits.
\newblock Schizophrenia research. 2019;206:118--126.

\bibitem{driver2013chi}
Driver DI, Gogtay N, Rapoport JL.
\newblock Childhood onset schizophrenia and early onset schizophrenia spectrum
  disorders.
\newblock Child and Adolescent Psychiatric Clinics. 2013;22(4):539--555.

\bibitem{sinclair2011dyn}
Sinclair D, Webster M, Wong J, Weickert C.
\newblock Dynamic molecular and anatomical changes in the glucocorticoid
  receptor in human cortical development.
\newblock Molecular Psychiatry. 2011;16(5):504--516.

\bibitem{markham2012sex}
Markham JA.
\newblock Sex steroids and schizophrenia.
\newblock Reviews in Endocrine and Metabolic Disorders. 2012;13:187--207.

\bibitem{gorzalka2012min}
Gorzalka BB, Dang SS.
\newblock Minireview: Endocannabinoids and gonadal hormones: bidirectional
  interactions in physiology and behavior.
\newblock Endocrinology. 2012;153(3):1016--1024.

\bibitem{engler2017est}
Engler-Chiurazzi E, Brown C, Povroznik J, Simpkins J.
\newblock Estrogens as neuroprotectants: estrogenic actions in the context of
  cognitive aging and brain injury.
\newblock Progress in neurobiology. 2017;157:188--211.

\end{thebibliography}

\section*{List of Abbreviations}
Abbreviations introduced in this paper are marked by *** and boldface. 

\vspace{10pt}
2AG: 2-arachidonoylglycerol. 

ALA: alpha-linoleic acid. 

ARA: arachidonic acid. 

ACh: acetylcholine. 

AEA: N-arachidonoylethanolamine.

CB1: type 1 cannabinoid receptor. 



COX1, COX2: cyclooxygenase-1, -2. 

cPLA2: calcium-dependent PLA2. 

CRH (also known as CRF): corticotropin-releasing hormone (or factor). 

CSF: cerebrospinal fluid. 

D2, D2R: dopamine 2 receptor. 

DA: dopamine. 

DAG: diacylglycerol. 

DAGL: diacylglycerol lipase. 

DHA: docosahexaenoic acid. 

ECB: \ecb. 

{\bf ***E-PUFA: PUFA derivates stimulated by \ecell\ \catwo\ (e.g., PGs). }

FAAH: fatty acid amide hydrolase. 

FEP: first-episode psychosis. 

GABA: gamma-aminobutyric acid. 

GC: glucocorticoid. 

GIRK channel: G protein-coupled inward rectifying potassium (\kplus) channel. 

GPCR: G protein-coupled receptor. 

H2O2: hydrogen peroxide. 

IIN: inhibitory interneuron. 

iPLA2: calcium-\indep\ PLA2.  

{\bf ***I-PUFA: PUFA derivates stimulated by \icell\ \catwo\ (e.g., AEA, Lkts). }


LA: linoleic acid. 

{\bf ***Lkt: leukotriene. }

LOX: lipoxygenase. 

LSD: lysergic acid diethylamide. 

MAGL: monoacylglycerol lipase. 

NAPE: N-arachidonoyl phosphatidylethanolamine. 

NAPE-PLD: NAPE phospholipase D. 

NEP: norepinephrine. 

NMDAR: N-methyl-D-aspartate receptor. 

NSAIDs: non-steroidal anti-inflammatory drugs. 

om3: omega 3 (PUFA). 

om6: omega 6 (PUFA). 

{\bf ***\thn : PUFA theory of \sz. }

PC: phosphatidylcholine. 

PCP: phenylcyclohexyl piperidine. 

PE: phosphatidylethanolamine. 

PFC: prefrontal cortex. 

PG: prostaglandin. 

PI: phosphatidylinositol.

PLA2: phospholipase A2. 

PLC: phospholipase C. 

PUFA: polyunsaturated fatty acid. 

RA: rheumatoid arthritis. 

SER: serotonin (5-HT). 

SER2a: 5-TH 2a receptor. 

SPMs: specialized pro-resolving mediators. 

SREBP: sterol regulatory element-binding protein. 

SZ: \sz.

TD: tardive dyskinesia. 


TRPV1: transient receptor potential vanilloid 1 (channel). 

\end{document}